\documentclass[journal=jacsat,manuscript=article]{achemso}
\usepackage[version=3]{mhchem}

\usepackage[normalem]{ulem}
\usepackage[usenames]{color}
\usepackage{bm}
\usepackage[dvipsnames]{xcolor}
\usepackage[symbol]{footmisc}

\usepackage[utf8]{inputenc}
\usepackage{microtype}

\usepackage{xr}

\makeatletter
\newcommand*{\addFileDependency}[1]{
  \typeout{(#1)}
  \@addtofilelist{#1}
  \IfFileExists{#1}{}{\typeout{No file #1.}}
}
\makeatother

\newcommand*{\myexternaldocument}[1]{%
    \externaldocument{#1}%
    \addFileDependency{#1.tex}%
    \addFileDependency{#1.aux}%
}

\myexternaldocument{si}

\title{Structure, Organization and Heterogeneity of Water-Containing Deep Eutectic Solvents}

\author{Kai T\"opfer} 
\altaffiliation{These authors contributed equally}
\affiliation{Department of Chemistry, University
  of Basel, CH-4056 Basel, Switzerland}

\author{Andrea Pasti}
\altaffiliation{These authors contributed equally}
\affiliation{Department of Chemistry, University of
  Zurich, CH-8057 Z\"urich, Switzerland}

\author{Anuradha Das} 
\altaffiliation{These authors contributed equally}
\affiliation{Institute of Applied Physics,
  University of Bern, CH-3012 Bern, Switzerland}

\author{Seyedeh Maryam Salehi}
\affiliation{Department of Chemistry, University
  of Basel, CH-4056 Basel, Switzerland}

\author{Luis Itza Vazquez-Salazar}
\affiliation{Department of Chemistry, University
  of Basel, CH-4056 Basel, Switzerland}

\author{David Rohrbach} 
\affiliation{Institute of Applied Physics,
  University of Bern, CH-3012 Bern, Switzerland}

\author{Thomas Feurer} 
\affiliation{Institute of Applied Physics,
  University of Bern, CH-3012 Bern, Switzerland}

\author{Peter Hamm} 
\affiliation{Department of Chemistry, University of
  Zurich, CH-8057 Z\"urich, Switzerland}

\author{Markus Meuwly} 
\affiliation{Department of Chemistry, University of
  Basel, CH-4056 Basel, Switzerland}
\email{m.meuwly@unibas.ch}

\begin{document}

\begin{tocentry}
\includegraphics[height=4.5cm]{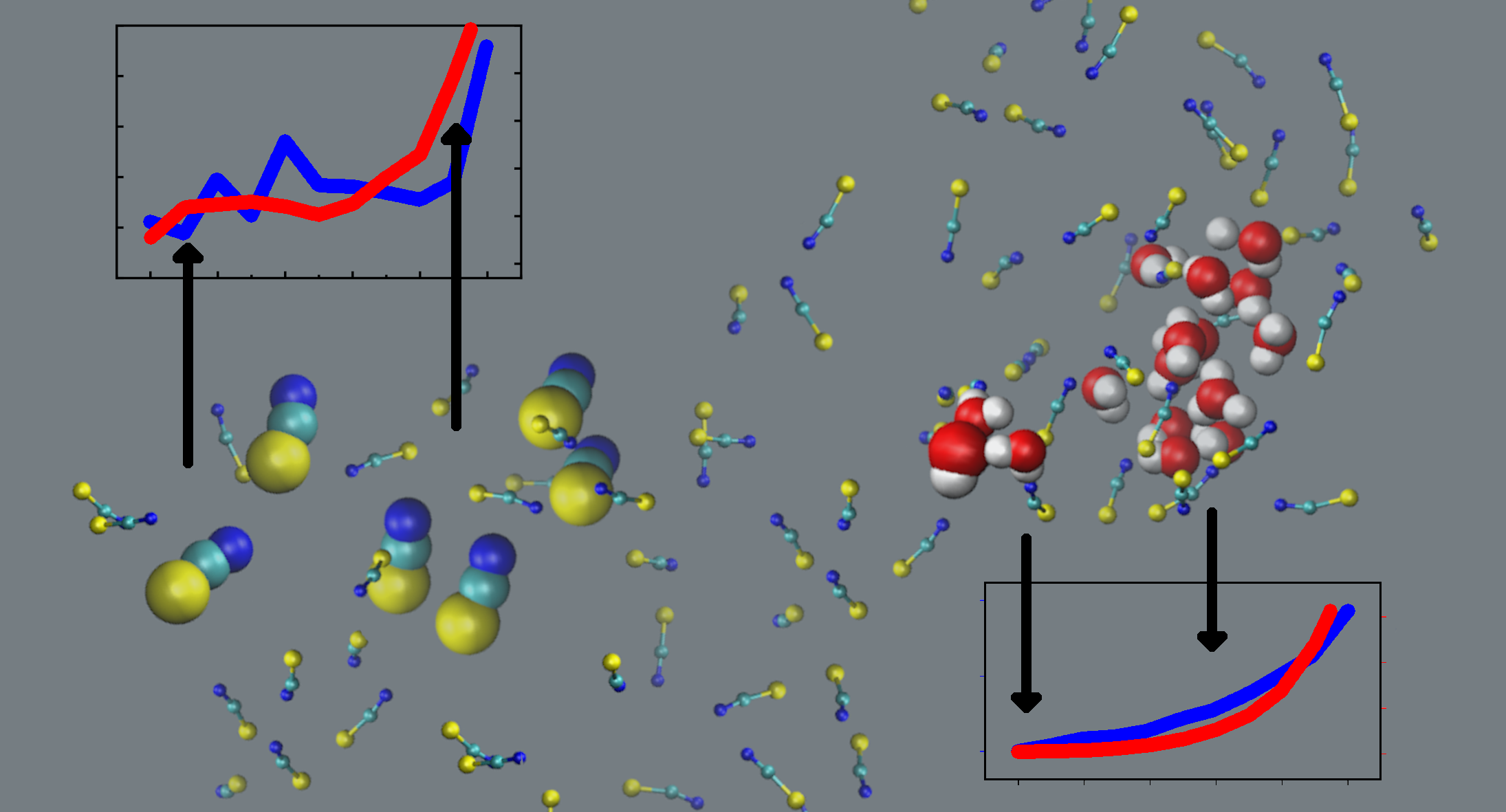} Deep eutectic
KSCN/acetamide mixtures with varying water content are investigated
experimentally by 2D infrared and terahertz spectroscopy and by
molecular dynamics simulations to characterize the heterogeneity of
such mixtures by modulating their composition.
\end{tocentry}

\begin{abstract}
The spectroscopy and structural dynamics of a deep eutectic mixture
(KSCN/ acetamide) with varying water content is investigated from 2D~IR
(with the C--N stretch vibration of the SCN$^-$ anions as the reporter)
and THz spectroscopy. Molecular dynamics simulations correctly
describe the non-trivial dependence of both spectroscopic signatures
depending on water content. For the 2D~IR spectra, the MD simulations
relate the steep increase in the cross relaxation rate at high water
content to parallel alignment of packed SCN$^-$ anions. Conversely,
the non-linear increase of the THz absorption with increasing water
content is mainly attributed to the formation of larger water
clusters. The results demonstrate that a combination of structure
sensitive spectroscopies and molecular dynamics simulations provides
molecular-level insights into emergence of heterogeneity of such
mixtures by modulating their composition.
\end{abstract}

\maketitle

\section{Introduction}
Deep eutectic solvents (DESs) are multicomponent mixtures
prepared by combining hydrogen bond acceptors (HBAs) and hydrogen
bond donors (HBDs), sometimes along with electrolytes at a
particular eutectic molar
ratio.\cite{abbott2003DES,marcus2019trends,martins2019defdes,hansen:2020}
These systems are characterized by a pronounced depression of their
melting point compared to their components and remain in the liquid
phase over a wider temperature interval.\cite{Smith2014} It is
believed that the high entropy gain in liquid range is supported by
strong H-bonding networks permeating the system, besides other
specific interactions of the various components in DES. DESs gained
considerable attention as potential substitutes for room temperature
ionic liquids due to their non-toxicity.\cite{Lomba2021} In addition
low vapor pressure, low melting point, high thermal stability, and
cheap production costs are some of the beneficial properties that
make these solutions interesting in chemical, biological, and
industrial
applications.\cite{emami2020desdrugreview,misan2019desfoodreview,majid2020desolireview,
cai2019applicationdesreview,marcus2019applicationsdesreview,
perna2019desgreenreview}\\

\noindent
For further improvement and optimization of such liquids, a more
molecularly resolved, microscopic understanding of their structural
and dynamical properties is required. DESs have been studied with a
range of approaches, including
dielectric,\cite{amico1987dielectric,berchiesi1999structural,berchiesi1992high,spittle:2022}
viscoelastic,\cite{berchiesi1995cryoscopic,berchiesi1983viscoelastic}
nuclear magnetic \cite{berchiesi1995cryoscopic,spittle:2022} and
ultrasonic
relaxation,\cite{berchiesi1995cryoscopic,farrat1992ultrasonic} optical
Kerr effect spectroscopy,\cite{biswas2014rikes} fluorescence
measurements,\cite{gazi2011fluorescence,guchhait2010fluorescence,guchhait2012medium,guchhait2014interaction}
wide-angle and quasi-elastic neutron
scattering\cite{spittle:2022}, simulated X-ray scattering structure
function calculations\cite{Kumari2018,kaur:2019,kaur:2020}
empirical potential structure refinement (EPSR),\cite{Hammond2017} and
molecular dynamics (MD)
simulations.\cite{pal2011heterogeneity,spittle:2022,Kaur2016} For
instance, femtosecond Raman-induced Kerr-effect spectroscopy
measurements allowed, in some cases, direct observation of
intermolecular hydrogen-bonding.\cite{biswas2014rikes} An important
characteristic of DESs - ``dynamic heterogeneity'' - is suggested by
the fractional dependence of a solutes solvation $\langle \tau_{\rm s}
\rangle$ or rotation $\langle \tau_{\rm r} \rangle$ time on viscosity
$\eta$ and temperature $T$, which scales as $\langle \tau_{\rm s/r}
\rangle \propto (\eta / T)^p$ with an exponent $0.4 < p <
0.7$.\cite{dinda:2021} In contrast, conventional Stokes-Einstein and
Stokes-Einstein-Debye would predict $p=1$. Multi- or
stretched-exponential relaxation of dynamic structure factors observed
in MD simulations explain this viscosity dependence through the
existence of spatially varying relaxation rates, reinforcing the idea
of a high degree of solution heterogeneity and the formation of
domains.\cite{guchhait2014interaction} These domains were found to
exhibit either collective small amplitude rotation or orientational
jump motions followed by H-bond relaxation\cite{Mukherjee2015} similar
to what was found for liquid water\cite{laage:2006} or around hydrated
ions.\cite{MM.cn:2013} The formation of ion aggregates leading to
charged structures in equilibrium with the liquid amide has also been
suggested by dielectric relaxation
measurements.\cite{berchiesi1992high}\\

\noindent
Optical spectroscopy is a versatile and powerful method to interrogate
and characterize the structural dynamics of condensed phase
systems. With regard to the present problem, an extensive literature
exists on the vibrational energy exchange between ions in solution
(especially
water),\cite{kuo07,bian2011ionclustering,bian2011nonresonant,bian2012ion,bian2013cation,chen2012ultrafast,Chen2014b,Chen2014a,Fernandez-Teran2020a}
establishing two-dimensional infrared (2D~IR) spectroscopy as a tool
to investigate structural and dynamical properties of multi-component
liquids through vibrational energy transfer between some of its
components. For pairwise two molecules, the vibrational energy
transfer rate is expected to have a steep, i.e. inverse 6th-power,
distance dependence,\cite{Chen2014a} similar to NOESY in NMR
spectroscopy,\cite{Schirmer1970,Bell1970} or F\"orster energy transfer
between electronic chromophores.\cite{Stryer1967} The presence of
cross-peaks in a 2D spectrum has been directly related to the
formation of aggregated structures, like ion
clusters.\cite{bian2011ionclustering}\\ 

\noindent
Terahertz (THz) spectroscopy, on the other hand, is sensitive to the
more collective excitations in a sample involving particle motion
leading to structural relaxations.\cite{Cunsolo2015} Specifically,
hydrogen bond networks exhibit intermolecular vibrations in the THz
spectral range. In turn, changes in THz absorption often carry
information on structural modifications of the hydrogen bond network
in a sample. For instance, when bulk water molecules become bound,
such as on surfaces or in proteins, the THz absorption
changes.\cite{smolyanskaya2018} Similarly, DESs are expected to
exhibit specific THz signatures for the heterogeneous mixture, because
a hydrogen-bonded DES constituent in aggregated structures is closer
to its bulk form as compared to being embedded in a homogeneous
mixture.\\

\noindent
One of the great challenges is to relate spectroscopic responses to
underlying structural motifs, their dynamics, and their change in
time. A typical example is ligand recombination in
myoglobin\cite{anfinrud:2003} for which time resolved experiments were
combined with atomistic simulations to characterize complex,
heterogeneous systems at molecular level.\cite{MM.mbco:2004} In the
area of molecular and ionic liquids, neutron
scattering\cite{castner:2010} and X-ray\cite{castner:2011} techniques
were used for a structural characterization combined with molecular
dynamics simulations. The present work reports on the structural
dynamics of the (0.75 acetamide + 0.25 KSCN) DES and its aqueous
solutions by replacing acetamide with 0 to 100\% water. The physical
properties around the eutectic temperature of 298~K have been
studied to verify its liquidous range.\cite{Kalita1998,DT9880002701}
It has been found that addition of inorganic salts lowers the
melting point of acetamide containing mixtures and are extensively
used as reaction media and in thermal salt cells to produce
electricity.\cite{HU200428,Wallace1972,guchhait2010fluorescence,Kalita1998}
In general, DESs are hygroscopic, hence it is important to
understand how their microscopic properties change with increasing
water content.\cite{Palmelund2021} Also, adding water changes the
physical properties of a DES, such as its viscosity or density, by
up to one order of magnitude.\cite{DAI2015} Hence to use DESs as
effective solvents, it is important to know how water - which is
both a HBD as well as a HBA - alters the arrangements of the DES
components, in particular the spectroscopically responsive part
which is SCN$^-$ in the present case. Thiocyanate (SCN$^-$) is an
ideal spectroscopic probe as the CN-stretch vibration absorbs in an
otherwise empty region of the infrared region, and has also been
used in recent work on urea/choline chloride mixtures which probed
the effect of water addition.\cite{bagchi:2021}\\

\noindent
In the present work two complementary spectroscopic techniques - 2D~IR
spectroscopy and THz spectroscopy - are combined with atomistic
simulations to probe the structural dynamics of a multi-component
DES. The experiments are sensitive to the aggregation of the various
components and characterization of a heterogeneous condensed phase
systems on multiple length and time scales. Combining both
spectroscopies with MD simulations, which essentially quantitatively
reproduce the spectroscopic experiments, allows one to portray the
short- and long-range order in DESs to a degree that is largely
inaccessible to other, more traditionally employed methods such as
NMR or dielectric spectroscopy, neutron scattering, or fluorescence
measurements.\\

\section{Methods}
\label{methods}

\subsection{2D~IR Spectroscopy}
The 2D~IR spectrometer in pump-probe geometry used in this work is a
slight modification of our previously reported
setup.\cite{Helbing2011} In brief, the output of a 5~kHz Ti:Sapphire
amplifier producing $\approx$100~fs pulses was used to pump a
home-built mid-IR OPA,\cite{hamm2000noise} delivering $\approx 2
\mu$J, ca. 120~fs pulses centered around 2020~cm$^{-1}$. A small
fraction of the mid-IR light was split off with a BaF$_2$ wedge to be
used as the probe and reference beams. Absorptive 2D~IR spectra were
obtained by fast scanning the coherence delay between the two pump
pulses (up to 5~ps, revealing a $\omega_{1}$ resolution of $\approx
2$~cm$^{-1}$ after zero-padding) generated in a Mach-Zehnder
interferometer for a given population delay ($t_{2}$). The second
output of the Mach-Zehnder interferometer was measured with a
pyroelectric detector, revealing the information to phase the 2D~IR
spectra.\cite{Helbing2011} The pump and probe beams were overlapped at
the sample position. Afterwards, the probe and reference beams were
dispersed in a spectrograph with a 150~l/mm grating, and
simultaneously recorded with a $2 \times 32$ pixels MCT detector,
yielding a $\omega_{3}$ resolution of $\approx 5$~cm$^{-1}$. Scatter
suppression was achieved by quasi-phase cycling using a librating ZnSe
window introduced in the pump beam at Brewster angle.\cite{bloem10}
2D~IR spectra were measured in transmission with pump and probe pulses
polarized in parallel.\\

\subsection{THz Spectroscopy}
The THz spectrometer is home-build and covers the frequency range
between 3~cm$^{-1}$ (0.1~THz) and 66 cm$^{-1}$ (2~THz). Terahertz
pulse generation and detection is based on a femtosecond erbium fiber
laser providing pulses with a duration of 100~fs, a repetition rate of
80~MHz, and an average output power of 130~mW at around 780~nm. The
laser pulses are split and sent to a photo-conductive emitter and
detector. For lock-in detection, the bias voltage of the emitter is
modulated with 2~kHz. The THz source is imaged by four aspheric lenses
with $f_1 = 100$~mm, $f_2 = 50$~mm, $f_3 = 50$~mm and $f_4 = 100$~mm,
each with a diameter of two inches, to the detection unit.  The
cuvette containing the different DES/water mixtures was located at the
intermediate image position. The transmitted THz electric field during
6~ps was measured in delay steps of 33~fs.  A reference signal (i.e.,
without sample) was measured before and after every DES/water
mixture. \\

\noindent
Terahertz spectra were obtained by Fourier transforming the measured
time dependent electric field. The spectrally resolved THz absorption
was calculated from the ratio of a sample and a reference spectrum and
the mean absorption is the absorption averaged between 13 cm$^{-1}$
(0.4~THz) and 34 cm$^{-1}$ (1~THz). The relative error was determined
as 2\%.\\

\subsection{Sample Preparation}
\noindent
For the 2D~IR experiments the salt was first prepared in a 50\%/50\% 
mixture of the two
isotopologues, KS$^{12}$C$^{14}$N and KS$^{13}$C$^{15}$N. Solutions
were prepared with constant total salt concentration of 4.4~M, and the
molar ratio of D$_2$O vs. acetamide varied from 0\% to 100\% in steps
of 10\%. The sample was squeezed between two CaF$_2$ windows without
any spacer, revealing a sample thickness of about 1~$\mu$m.\\

\noindent
For the THz measurements the same preparation recipe was applied, but
using H$_2$O instead of D$_2$O and without specific measures to
control the relative isotope concentration. The DES/water mixtures
were filled in a 100~$\mu$m thick fused silica cuvette with a 1.3~mm
thick base plate and a 1.2~mm thick cover plate. All samples were
measured at a temperature of 295~K.\\

\subsection{MD Simulations}
Molecular dynamics (MD) simulations were performed using
CHARMM\cite{Charmm-Brooks-2009} with the all atom force field
CHARMM36.\cite{CHARMMFF-ALL36-Guvench2011} The intramolecular and
intermolecular potential of acetamide, water and KS$^{12}$C$^{14}$N
are determined by the bond parameters of CGenFF\cite{CGenFF} and the
TIP3P model \cite{TIP3P-Jorgensen-1983} for water because CGenFF was
parametrized at 300 K with this water model. The CN and SC bond
potential in SCN$^-$ are replaced by a Morse potential and
electrostatic multipole moments of SCN$^-$ up to the quadruple moment
are included in the force field. The electrostatic
moments\cite{MM.mtp:2013,mm.mtp2:2016,mm.mtp:2017} and Morse
parameters are fitted to the respective \emph{ab initio} data for a
SCN$^-$ anion at MP2/aug-cc-pVDZ level of theory using
Gaussian09.\cite{gaussian09} All force field parameters used for the
present study are summarized in the supporting information, see Table
S1. Bonds involving hydrogen atoms were constrained
using the SHAKE algorithm.\cite{shake77} For the nonbonded
interactions the long-range cutoff was 14~\AA\/ and electrostatic
interactions were controlled using the Particle Mesh Ewald
algorithm.\cite{Darden1993} The molar composition of the systems was
changed by modifying the number of water and acetamide molecules while
keeping constant the total concentration of KSCN. The number of
molecules and molar fractions are reported in Table
S2. In total, 10 independent random initial
configurations for each of the 12 system compositions were set up
using PACKMOL.\cite{martinez:2009} Following 180~ps of equilibrium,
dynamics simulation in the \emph{NpT} ensemble at 300~K and normal
pressure, production simulations were performed for 5~ns with a time
step of 1~fs using the leap-frog integrator and a Hoover
thermostat.\cite{Hoover1985} In total, $50$~ns for each system
composition were sampled. The densities of the different system
compositions is reported in the supporting information, see Figure
S1 together with results from using alternative force field
parametrizations.\\

\noindent
For computing the 2D IR spectra the frequency trajectories
$\omega_i(t)$ for each oscillator are required. These were
determined from an instantaneous normal mode
(INM)\cite{stratt:1994, keyes:1997, schmuttenmaer:1997, saito:2013}
analysis for the CN vibrational frequencies $\omega_i$ on snapshots of
the trajectories each $200$~fs in $250$~ps time interval at
$0-0.25$~ns, $2-2.25$~ns and $4-4.25$~ns on all SCN$^-$ ions.
For this the structure of every SCN$^-$ ion was optimized while
freezing the position of all remaining atoms in the system, followed
by a normal mode analysis using the same force field that was
employed for the MD simulations. Previously, such an approach has
been validated for N$_3^-$ in solution by comparing with rigorous
quantum bound state calculations.\cite{MM.n3:2019}\\

\subsection{Simulation and Analysis of 2D~IR Spectra}
To calculate 2D~IR spectra from the output of the MD simulations, we
started from the exciton Hamiltonian for $N$
oscillators:\cite{hamm11,Liang2012,Fernandez-Teran2020a}
\begin{align}\label{eq:ExcitonHamiltonian}
	\bm{H}(t) =&
        \sum_{i=1}^{N}\left[\omega_{i}(t)\bm{b}_{i}^{\dagger}
          \bm{b}_{i} -
          \frac{\Delta}{2}\bm{b}_{i}^{\dagger}\bm{b}_{i}^{\dagger}\bm{b}_{i}\bm{b}_{i}\right]
        + \sum_{i\neq j}^{N} V_{ij}(t)\bm{b}_{i}^{\dagger}\bm{b}_{j}
        \\ \nonumber + &\sum_{i=1}^{N}
        \vec{\mu}_{i}(t)\cdot\vec{E}(t)\left[\bm{b}_{i}^{\dagger}+\bm{b}_{i}\right],
\end{align}
with
\begin{equation}\label{eq:BetaCoupling}
	V_{ij} =
        \frac{1}{4\pi\varepsilon_{0}}\left[\frac{\vec{\mu}_{i}\cdot\vec{\mu}_{j}}{r_{ij}^{3}}-3\frac{(\vec{r}_{ij}\cdot\vec{\mu}_{i})(\vec{r}_{ij}\cdot\vec{\mu}_{j})}{r_{ij}^{5}}
          \right].
\end{equation}
Here, $\bm{b}_{i}^{\dagger}$ and $\bm{b}_{i}$ are the harmonic
creation and annihilation operators of a vibrational quantum on
SCN$^-$ anion $i$, $\omega_i(t)$, $\vec{\mu}_{i}(t)$ and
$\Delta=27~\rm{cm}^{-1}$ its instantaneous frequency, transition
dipole and anharmonicity, respectively, $r_{ij}$ the distance between
both sites, $\varepsilon_0$ the vacuum permittivity, and $\vec{E}(t)$
the applied external field from the laser pulses. The instantaneous
frequencies $\omega_i(t)$, directions of the transition dipoles
$\vec{\mu}_{i}(t)$ and connecting vectors $\vec r_{ij}(t)$ are
time-dependent and can be extracted directly from the MD
trajectory. To that end, the directions of the transition dipoles
$\vec{\mu}_{i}(t)$ were assumed to be along the CN bond of each
SCN$^-$ ion, its origin in the middle of the CN bond, and the strength
0.35~D,\cite{Chen2014b} while the frequencies $\omega_i(t)$ were
derived from INM calculations (see above). \\

\noindent
2D~IR spectra were calculated from this Hamiltonian by formulating the
6 response functions of ground state bleach (GB), stimulated emission
(SE) and excited state absorption (EA) in the two phase matching
directions that contribute to purely absorptive 2D~IR spectra, see
Eq.~15 of Ref.~\citenum{Liang2012}. As in the experiments, the
polarizations of all field interactions was set parallel to each
other, and $\left(\langle xxxx\rangle+\langle yyyy\rangle+\langle
zzzz\rangle\right)/3$ has been calculated to better average over the
orientations of the SCN$^-$ anions. For the time-evolution of the
wave functions, however, a somewhat different approach was taken than
in Ref.~\citenum{Liang2012}. In brief, the Trotter expansion in
Ref.~\citenum{Liang2012} was replaced by the Chebyshev
method,\cite{Tal-Ezer1984} which allows for larger time-steps and thus
speeds up the calculation significantly, as discussed in
Ref.~\citenum{Fernandez-Teran2020a}. Since the computational cost
scales with the inverse 3rd power of the time step in the
time-propagation of the wavefunction, a relatively long time step of
200~fs was chosen. This is longer than the decay time of the initial
spike in the FFCF (see Figure S2, Supplementary
Material), and thus overestimates the homogeneous width of the
simulated spectra. However, for the present analysis only the total
peak volumes are required as a measure for the overall
populations. Inhomogeneity causes the peaks to be tilted, expressed
by the correlation parameter $c$ in Eq.~\ref{sieq:2DGC} below, and
is accounted for in the calculation of the total peak volume, see
Eq.~\ref{sieq:2DGCvol}. Working in a rotating frame, 2D~IR spectra
were calculated with coherence times of 3.2~ps and the population time
was varied in steps of 1~ps up to 100~ps.\\

\noindent
A sliding average along the 250~ps long sequences of the trajectories
was performed in steps of 5~ps in order to increase the
signal-to-noise ratio of the calculated 2D~IR spectra. For each sample
in this sliding average, a new 50\%/50\% distribution of the two
isotopologues of the SCN$^-$ anion was randomly selected in order to
generate samples as statistically independent as possible. The
instantaneous frequencies $\omega_i(t)$ of those anions, which were
selected to be KS$^{13}$C$^{15}$N, were lowered by $75~\rm{cm}^{-1}$
to account for the isotope shift. The 2D~IR spectra for each
mixing ratio were obtained by averaging the $30$ ($3$ time intervals
per 10 samples for each mixing ratio) 2D~IR spectra for each 250~ps
long fragments\\

\noindent
To extract the essential information from the 2D~IR spectra, i.e., the
frequency correlation of the diagonal peaks as well as the cross-peak
intensities, they were fit globally similar to
Ref.~\citenum{Fernandez-Teran2020a}. To that end, each 2D~IR peak
was modelled as a correlated 2D-Gaussian function:
\begin{align} \label{sieq:2DGC}
	G(\omega_{1},\omega_{3}) &= A \exp \biggl\{
        -\frac{1}{2(1-c^{2})}
        \biggl[\biggl(\frac{\omega_{1}-\omega_{1,0}}{\Delta{\omega}}\biggr)^{2}
          \\ +
          &\biggl(\frac{\omega_{3}-\omega_{3,0}}{\Delta\omega}\biggr)^{2}
          - \frac{
            2c(\omega_{1}-\omega_{1,0})(\omega_{3}-\omega_{3,0}) }{
            \Delta\omega^2} \biggr] \biggr\}, \nonumber
\end{align}
where $A$ is an amplitude, $\omega_{i,0}$ the center frequencies along
frequency axis $i$, $\Delta\omega$ the frequency width, and $c$ a
correlation coefficient as a measure of spectral
diffusion.\cite{Guo2015} In total 8 such peaks were defined for a
2D~IR spectrum, with many parameters set equal to minimize their total
number in the fitting. That is, two pairs of diagonal peaks were
defined for the two isotopologues with different central frequencies
$\omega_{i,0}$ and amplitudes $A$, but identical width $\Delta\omega$
and correlation $c$, since the latter two parameters describe the
identical interaction of the ions with the solvent environment. Each
diagonal peak consists of ground state bleach (GB)/stimulated emission
(SE) as well as excited state absorption (EA), where the latter has
been down-shifted in the $\omega_3$-direction by the anharmonicity
$\Delta=27~\rm{cm}^{-1}$. The amplitudes $A$ and correlations $c$ of
the two contributions to each diagonal peak were assumed to be the
same. In addition, two cross peak pairs were introduced, each again
consisting of GB/SE and EA contributions, for which center frequencies
and width were assumed to be the same as for the corresponding
diagonal peaks along the two frequency axes. The amplitudes $A$ of the
two cross peak pairs were set to be the same, and we assumed $c=0$ for
the correlation, since cross-relaxation randomizes the frequency. To
further stabilize the fitting procedure, the center frequencies
$\omega_{i,0}$ and the width $\Delta\omega$ of the two diagonal peaks
were first determined from the earliest 2D~IR spectrum at
$t_2=0.25$~ps, and then kept as fixed parameter when fitting a
population time series of 2D~IR spectra.\\

\noindent
The population is proportional to the peak volume
$\Phi$,\cite{Kwak2006} which was calculated from the obtained fitting
parameters according to
\begin{equation}
  \label{sieq:2DGCvol}
	\Phi = 2\pi A \Delta\omega^2 \sqrt{1-c^{2}}.
\end{equation}
The intensity of a cross peak scales with $\left\lVert
\vec{\mu}_{\rm{pu}} \right\rVert^{2} \left\lVert \vec{\mu}_{\rm{pr}}
\right\rVert^{2}$, where $\rm{pu}$ and $\rm{pr}$ stand for the pumped
and probed oscillators, respectively. With this in mind, the energy
transfer kinetics can then be retrieved by normalizing the cross peak
volumes with $\sqrt{\Phi_{\rm{pu}} \Phi_{\rm{pr}}}$, where
$\Phi_{\rm{pu}}$ and $\Phi_{\rm{pr}}$, respectively, are the volumes
of the corresponding diagonal peaks at early population times. The
fitting was performed in the same way for both experimental and
simulated spectra.\\

\subsection{Computation of the THz Spectra}
The THz absorption spectra are determined from the MD simulations
by evaluating the Fourier transform of the $\left<
\dot{\mu}(t)\dot{\mu}(0) \right>$ auto-correlation
function\cite{ramrez04}
\begin{equation}
  I_\mathrm{THz}(\omega) n(\omega) \propto \mathrm{Im}\int_0^\infty
  dt\, e^{i\omega t} \left \langle \boldsymbol{\dot \mu}_{\rm tot}(t)
  \cdot {\boldsymbol{\dot \mu}_{\rm tot}}(0) \right \rangle
\label{eq:THz}
\end{equation}
Here, $\boldsymbol{\dot \mu}$ is the time derivative of the dipole
moment $\boldsymbol{\mu}$ of the entire simulation system with
periodic boundary conditions correctly resolved for molecules that
cross boundaries. Partial THz absorption spectra were computed by the
Fourier transform from the auto-correlation $\left< \boldsymbol{\dot
  \mu}_{\alpha}(t)\boldsymbol{\dot{\mu}}_{\alpha}(0) \right>$ where
$\alpha$ denotes a respective species $\alpha=$\{SCN$^-$, K$^+$,
H$_2$O, acetamide\} of the system. The Fourier transform of the
cross-correlation $\left< \boldsymbol{\dot
  \mu}_{\alpha}(t)\boldsymbol{\dot{\mu}}_{\beta}(0) \right>$ between
species $\alpha$ and $\beta$ determines the contribution of the two
species due to their interplay to the absorption intensity.
Figure~S3 shows partial THz absorption spectra and the
Fourier transform of cross-correlation for a selected species and
species combinations, respectively. The integrated absorption or cross
correlation is the integral of the experimental or computed THz
absorption spectra, the computed partial absorption or the cross
correlation intensities in the range from 13 to 34~cm$^{-1}$.\\

\section{Results}
The sample for the present work is a multi-component DES system, which
allows for continuous tuning of the hydration state and consequently
the viscosity. That is, the eutectic system acetamide/KSCN is mixed
with water, which is especially well suited for the purpose of this
work, since the CN vibration in the thiocyanate anion (SCN$^-$) is an
ideal molecular probe for 2D~IR spectroscopy. Moreover the expected
hydrogen bonding water network in this mixture can be readily probed
via THz spectroscopy. Similar eutectic mixtures, i.e., acetamide,
potassium thiocyanate together with ethanol have been characterized
experimentally using thermodynamic
measurements\cite{liu2016properties} in view of their industrial
applications and the acetamide-KSCN mixture with water has been
  reported to remain liquid down to 303 K.\cite{liu:2013}\\

\begin{figure}
\centering \includegraphics[width=0.75\textwidth]{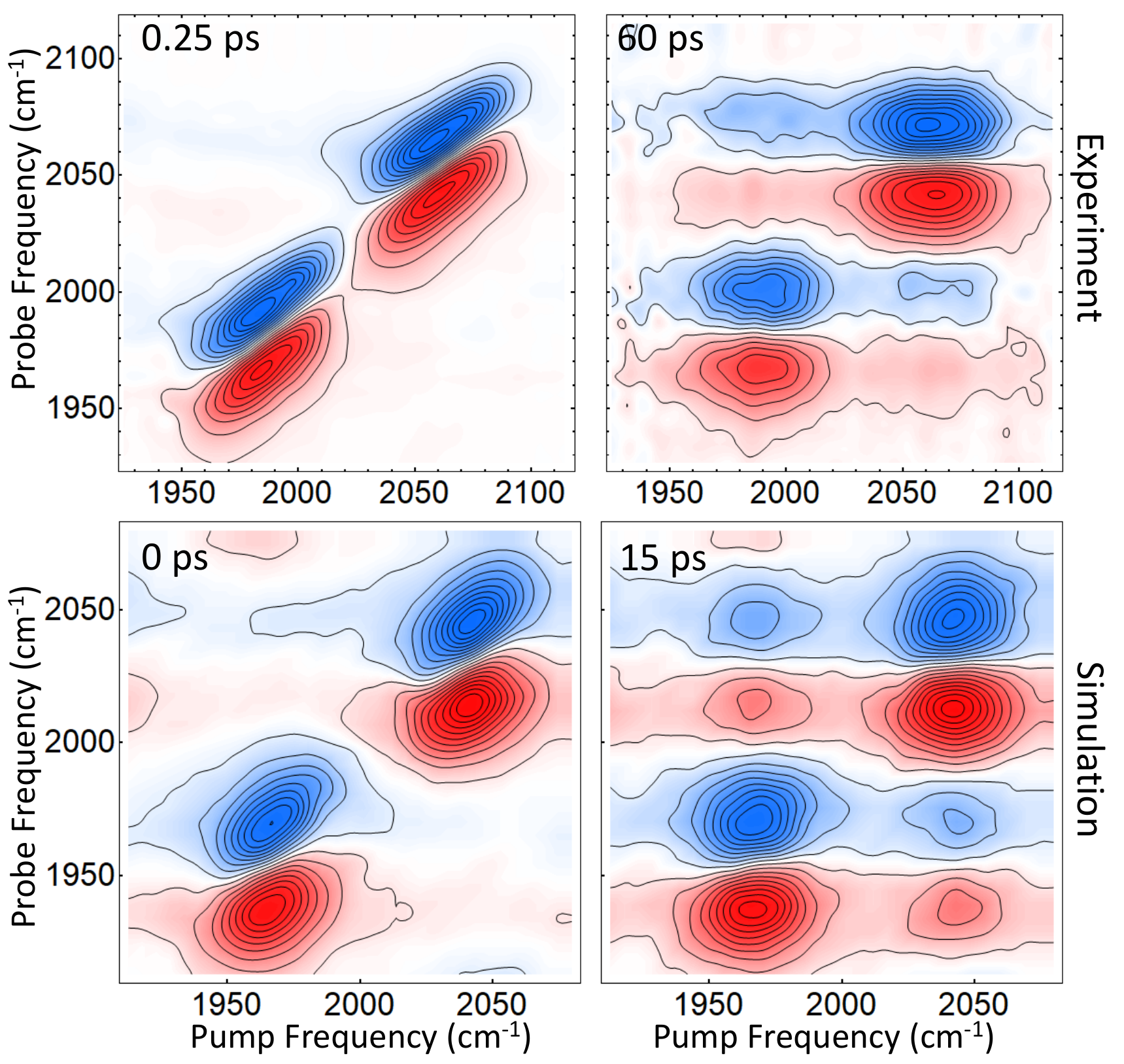}
\caption{Experimental (top) and simulated (bottom) 2D~IR spectra at
  early (left) and late (right) populations times (which are indicated
  in the panels), exemplified here for the sample with a 50\%/50\%
  mixture of D$_2$O/acetamide. The contour lines have been normalized
  to the peak of each spectrum, thereby suppressing $T_1$-relaxation.}
\label{fig:1}
\end{figure}

\noindent
For the 2D~IR experiments the salt was first prepared in a 50\%/50\%
mixture of the two isotopologues, KS$^{12}$C$^{14}$N and
KS$^{13}$C$^{15}$N. The two isotopologues are needed in the 2D~IR
experiment in order to have two resolved vibrational bands, between
which vibrational energy transfer can be observed. Figure~\ref{fig:1},
top, shows selected 2D~IR spectra at early (left) and late (right)
population times for the sample with a 50/50\% molar ratio of
D$_2$O/acetamide. At early population times (left) in essence only
diagonal peaks arise. They are tilted along the diagonal due to a
correlation between pump- and probe frequency, reflecting the spatial
inhomogeneity of the sample.\cite{hamm11} At later population times
(right), this correlation is lost (no tilt along the diagonal) but
additional cross peaks arise due to cross-relaxation between the
resonances of the two isotopologues KS$^{12}$C$^{14}$N and
KS$^{13}$C$^{15}$N. For such a cross peak to appear, one of the
isotopologues is excited at its vibrational frequency, its vibrational
energy is transferred to the other isotopologue during the population
time, and is then probed at the vibrational frequency of the second
isotopologue.\\ 

\noindent
The vibrational energy transfer rate is expected to have a steep,
i.e. inverse 6th-power, distance dependence for pairwise two
molecules. To that end, an analytical expression has been
derived\cite{Chen2014a}
\begin{equation}\label{eq:VET_rate}
  k = \frac{2}{1+\exp\left({-\frac{\hbar\Delta\omega}{k_{\textrm{B}}T}}\right)}
  \left[\frac{V^{2}~\tau^{-1}}{(\Delta\omega)^2 + 4V^2 + \tau^{-2}}\right],
\end{equation}
with $\Delta\omega$ the frequency difference between both molecules,
$\tau$ a dephasing correlation time, and the transition dipole
coupling $V$ taken from Eq.~\ref{eq:BetaCoupling}. The physical
picture behind this model is the following: Due to the coupling $V$,
excitation energy is initially transferred from the donor to the
acceptor in a coherent sense, as described by the time-propagation of
the time-dependent Schr\"odinger equation. The coherence between donor
and acceptor state is then abruptly terminated by a random dephasing
event after an average time $\tau$,\cite{Chen2014a} revealing a
first-order kinetics with rate described by Eq.~\ref{eq:VET_rate}. The
length scale over which vibrational energy transfer can be observed,
i.e., the equivalent of a ``F\"orster radius'', is 4-5~\AA~ in the
very best case, depending on the strength of the transition dipoles
and the lifetime of the vibrational excitation. Histograms for
vibrational coupling strengths $V$ for the different mixtures are reported
in Figure S4. Previously, it has been verified by numerical
propagation of the Schr\"odinger equation that the analytical model
Eq.~\ref{eq:VET_rate} is valid for dimers.\cite{Fernandez-Teran2020a}
It has, however, also been shown that the situation is more complex
when networks of interacting vibrational transitions exist, with a
percolation threshold at a certain level of
dilution.\cite{Fernandez-Teran2020a} Therefore, the fully coupled
systems are considered here.\\

\noindent 
The starting point for calculating 2D IR spectra are all-atom MD
simulations with the same relative concentrations of the various
components as in the experiments and a multipolar representation of
the electrostatics for the SCN$^-$ ions.\cite{MM.mtp:2013} The
instantaneous frequencies $\omega_i(t)$, the couplings $V_{ij}(t)$
between oscillators $i$ and $j$ and accompanying parameters are all
determined from the MD trajectories as input for the time-dependent
Frenkel exciton Hamiltonian Eq.~\ref{eq:ExcitonHamiltonian}. Computed
2D~IR spectra are reported in (Figure~\ref{fig:1}, bottom) and agree
with the experimental ones (Figure~\ref{fig:1}, top) extremely well.
Typical frequency-frequency correlation functions (FFCFs)
determined from the INM frequencies are reported in Figure
S2.\\

\noindent
For this comparison it is important to stress that the force field
used in the MD simulations has been derived independently and has not
been adapted to reproduce the experimental 2D results.  Only one
additional parameter was added for the calculation of the 2D~IR
spectra: the modulus of the transition dipole $\mu$ of the SCN$^-$
oscillators. The value used here was 0.35~D, which has been determined
experimentally before.\cite{Chen2014b} Additional electronic structure
calculations confirmed this value ($\sim 0.33$~D), but also showed
that considerable charge fluxes on the SCN$^-$ ion are responsible for
its magnitude. This effect is not accounted for in the force field
used here. Overall, cross relaxation is faster in the simulation,
which can be seen from the ratio of cross vs. diagonal peaks (also note
the different population times in Figure~\ref{fig:1}, right: 60~ps for
the experimental vs. 15~ps for the simulation spectra). Since the cross
relaxation rate is expected to scale with the 4th power of the
transition dipole moment,\cite{Chen2014a} it is concluded that the
assumed value overestimates the true transition dipole strength by
only a modest factor. Furthermore, Eq.~\ref{eq:VET_rate} suggests
that the cross-relaxation rate scales as $1 / \tau$ of the FFCF, and
therefore the long time step in the simulation (see Methods)
underestimates the cross-relaxation rate.\\

\begin{figure}
\centering
\includegraphics[width=0.75\textwidth]{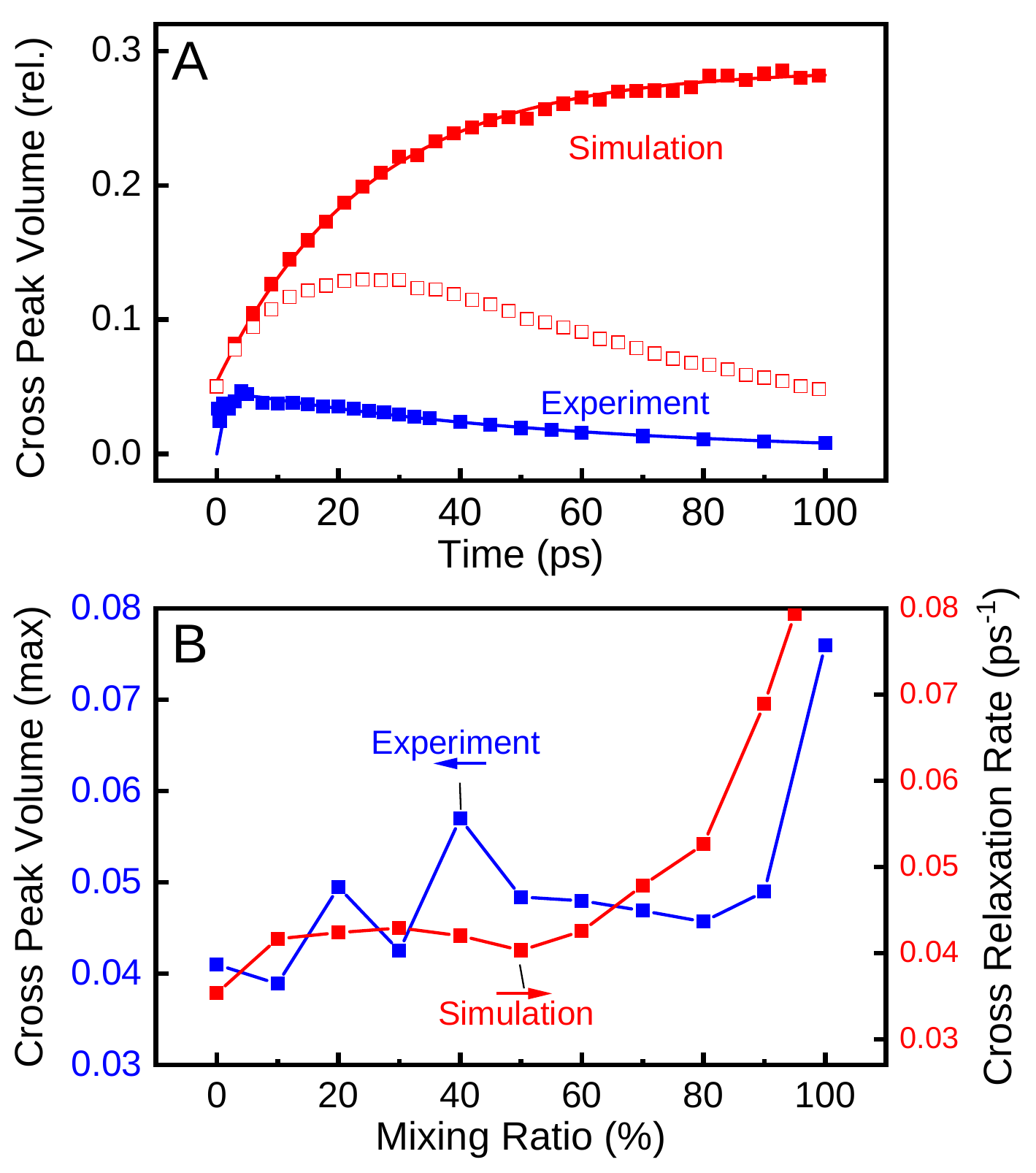}
\caption{(A) Cross peak volumes of the experimental (filled blue
  squares) and simulated (filled red squares) 2D~IR spectra
  for the sample with a 50\%/50\% mixture of D$_2$O/acetamide. The
  cross peak volumes have been normalized to the diagonal peak volumes
  at the earliest population times. The open red squares show the
  simulated data after phenomenologically multiplying them with a
  $T_1$ lifetime of 56~ps. The solid lines show the respective
  fits. (B) Comparison of the cross-relaxation rate as a function
  of mixing ratio. For the simulated 2D~IR spectra, the inverse time
  constants from single exponential fits are shown (right scale); in
  the case of the experimental 2D~IR spectra the peak amplitude $a_0$
  (left scale). Both data sets are scaled relative to each other.}
\label{fig:2}
\end{figure}

\noindent
The cross peak volumes as determined from Eq.~\ref{sieq:2DGC} from the
experimental (blue) and simulated (red) 2D~IR spectra as a function of
time for the sample with a 50\%/50\% mixture of D$_2$O/acetamide are
shown in Figure~\ref{fig:2}A. For the simulated spectra the cross-peak
volume exponentially approaches an equilibrium and the
cross-relaxation rate can be obtained by single-exponential fitting of
this data (revealing a time constant of 25~ps for the example shown in
Figure~\ref{fig:2}A). In contrast, cross relaxation in the experiments
competes with $T_1$ relaxation, an effect that is not included in the
exciton Hamiltonian Eq.~\ref{eq:ExcitonHamiltonian}. The cross peak
volume therefore rises initially and decays again later due to $T_1$
relaxation, see Figure~\ref{fig:2}A, blue. These data are fit with a
bi-exponential function, $-a_0(e^{-t/\tau_1}-e^{-t/\tau_2})$,
revealing time constants of 1.4~ps and 56~ps for the example shown in
Figure~\ref{fig:2}A. Since $\tau_1 \ll \tau_2$, the amplitude $a_0$
reflects in essence the peak of the signal, and serves as a measure of
the efficiency of cross relaxation. The open red squares in
Figure~\ref{fig:2}A show the simulated data after phenomenologically
multiplying them with a $T_1$ lifetime of 56~ps. The data is
qualitatively the same as the experimental ones, but the peak
amplitude is higher since the simulation overestimates the transition
dipole strength (as discussed above).\\

\noindent
Figure~\ref{fig:2}B shows the results of this analysis as a function
of acetamide/water mixing ratio, i.e., the single exponential cross
relaxation rate for the simulated 2D~IR spectra, and the amplitude
$a_0$ for the experimental 2D~IR spectra. While these two quantities
cannot be compared one-to-one, when properly scaling them relative to
each other, one can see very good agreement. In both cases, cross
relaxation varies non-monotonically as a function of the water
content, with a dip after an initial rise and a steep increase towards
higher water content.  Since the cross-relaxation rate is a steep
function of the intermolecular distances between the SCN$^-$ anions,
the cross relaxation rate reflects the clustering of the SCN$^-$
anions.  Overall, the theoretical curve is shifted somewhat towards a
lower water content as compared to the experimental curve, indicating
that the MD force field slightly overestimates SCN$^-$ clustering. A
simulation with 100\% water content no longer gave reasonable results
due to ion aggregation, and 95\% water content is the highest value
considered in all subsequent simulation results. Nonetheless, the fact
that the simulations are able to qualitatively capture this
non-trivial dependence of the cross relaxation rate on the mixing
ratio is remarkable.\\

\begin{figure}
\centering \includegraphics[width=0.99\textwidth]{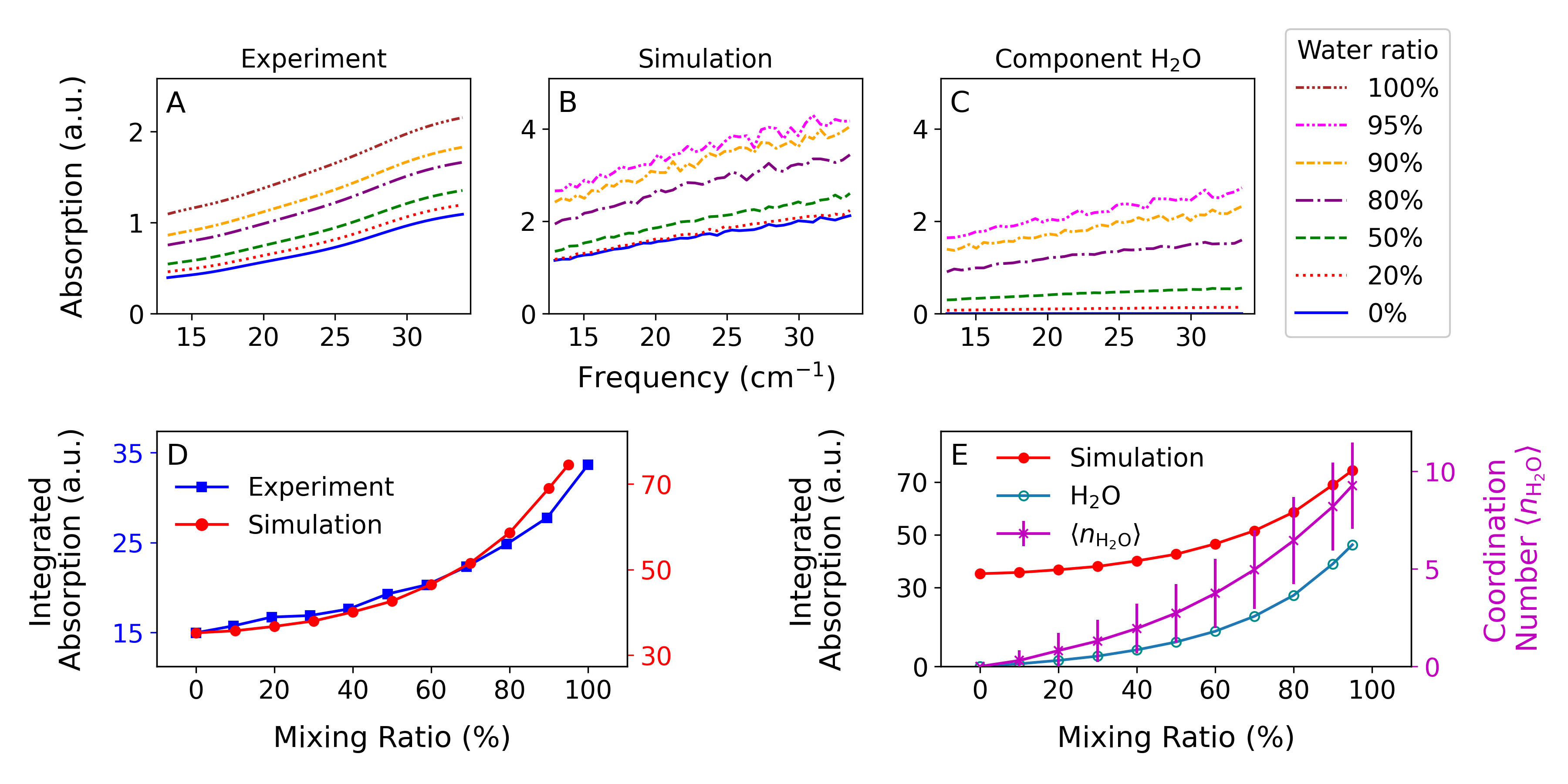}
\caption{Experimental (A) and simulated THz spectra (B) between
  13~cm$^{-1}$ to 34~cm$^{-1}$. Results are shown for different
  water/acetamide mixing ratios, including 0\% (blue), 20\% (red),
  50\% (green), 80\%(violet), 90\% (orange), 95\% (magenta) and 100\%
  (brown).  Panel C shows the THz absorption spectra from all H$_2$O
  molecules. Panel D compares the integrated intensity of the
  experimental (blue squares) and simulated (red circles) THz spectra
  between 13~cm$^{-1}$ and 34~cm$^{-1}$ for different water/acetamide
  mixing ratios. The data are shifted and scaled to best overlap the
  lowest and highest integrated absorptions from experiments and
  simulations. Panel E compares the integrated absorption intensity of
  the simulated THz spectra with the contribution from the water-water
  autocorrelation function (light blue). The magenta line with marker
  shows the average coordination number $\langle n_\mathrm{H_2O}
  \rangle$, which corresponds to the number of water molecules
  surrounding a central water molecule in mixtures with different
  mixing ratios. Water molecules are counted as in vicinity if the
  oxygen-oxygen distance is shorter than 4.5~\AA. The vertical lines
  show the standard deviation of the coordination number distribution
  $P(\langle n_\mathrm{H_2O} \rangle)$.}
\label{fig:3}
\end{figure}

\noindent
Measured Terahertz spectra are reported in Figure~\ref{fig:3}A for
different water/acetamide mixing ratios in the frequency range from
13~cm$^{-1}$ and 34~cm$^{-1}$. The samples were the same as for the
2D~IR experiment, except for the fact that H$_2$O and only naturally
abundant KS$^{12}$C$^{14}$N at 4.4~M concentration was used. The THz
absorption monotonically increases with wavenumber, consistent with a
broad band whose peak is outside of the measurement frequency
range. Its amplitude scales nonlinearly with mixing ratio across the
entire spectral range, which is seen best when plotting the absorption
integrated between 13~cm$^{-1}$ and 34~cm$^{-1}$ against the mixing
ratio, as shown in Figure~\ref{fig:3}D (blue squares). The nonlinear
increase is indicative of a non-trivial structural rearrangement with
increasing mixing ratio.\\

\noindent
The simulated THz absorption spectra (Figure~\ref{fig:3}B) agree well
with the measurements (Figure~\ref{fig:3}A) in that they reproduce the
increase in absorption with wavenumber and in particular the nonlinear
increase with mixing ratio. For a quantitative comparison, the
simulated integrated absorption was scaled such that it matches with
the measured results for its lowest and highest values, as shown in
Figure~\ref{fig:3}D, and the nonlinear increase for intermediate mxing
ratios is reproduced very well.\\

\noindent
Contributions from individual components to the total THz absorption
spectrum were determined from $\boldsymbol{\dot \mu}_{\rm
  tot}=\boldsymbol{\dot \mu}_{\rm water}+\boldsymbol{\dot \mu}_{\rm
  SCN^-}+\boldsymbol{\dot \mu}_{\rm K^+}+\boldsymbol{\dot \mu}_{\rm
  acetamide}$, where $\boldsymbol{\dot \mu}_{\alpha}$ is the
contribution of species $\alpha$ to the total dipole moment time
derivative of the simulation system (even for single charged atoms
there is a contribution to the total system dipole moment). This gives
rise to 4 auto-correlations and 6 cross-correlations when evaluated in
Eq.~\ref{eq:THz}. The largest contribution to the THz absorption
spectrum, as well as its nonlinear dependence on the mixing ratio, can
be traced back to the water-water autocorrelation function in the
mixture, see Figures~\ref{fig:3}C and the red vs. light blue line
in \ref{fig:3}E.  All other contributions to the total absorption
spectrum are comparably small, and are essentially flat as a function
of mixing ratio, see Figure~S3.\\

\noindent
To summarize this part, the MD simulations can almost quantitatively
reproduce the results of both the 2D~IR and THz spectroscopy, and in
particular capture correctly the highly non-trivial changes of the
spectroscopic responses for different mixing ratios. 2D~IR
spectroscopy is primarily sensitive to short range SCN$^-$/SCN$^{-}$
contacts and reveals more rapid intermolecular energy redistribution
between the CN stretch vibrations of SCN$^-$ in systems with higher
water ratios. The striking feature is the sharp increase of these
contacts at $\sim 90$\% water content and above which is also what
the simulations find but already at 80\% water content. On the other
hand, THz spectroscopy, which is mostly sensitive to the water in the
sample, shows a similar sharp increase in absorption above 80\% water
ratio. Both spectroscopic observations reflect and report on different
structural properties of the liquid. Based on this agreement between
experiments and simulations the structural properties, which can be
directly assessed from the MD simulations, will be analyzed in the
following.\\

\section{Discussion}
\begin{figure}
\centering \includegraphics[width=0.99\textwidth]{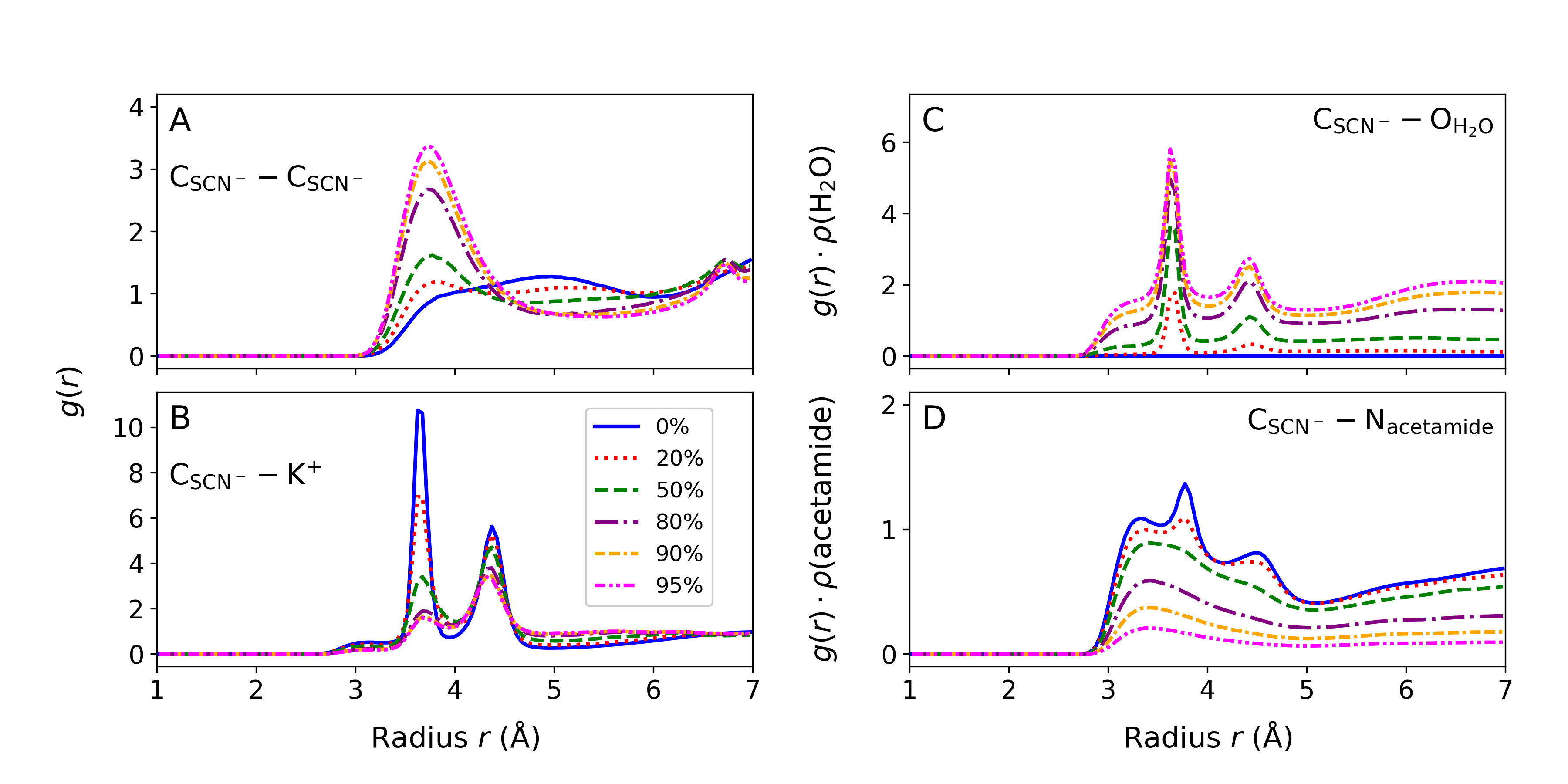}
\caption{Radial distribution function $g(r)$ for the C atom of SCN$^-$
  to (A) the C atom of SCN$^-$, (B) K$^+$, (C) the O atom of H$_2$O
  and (D) the N atoms of in acetamide (multiplied by its density
  $\rho$(H$_2$O) and $\rho$(acetamide), respectively) of the MD
  simulation of the different mixtures of H$_2$O/acetamide. See Figure
  S5 for $g_{\rm SCN^{-}-SCN^{-}}(r_{\rm CC})$ at the
  experimental density of 1.14 g/cm$^3$ for high water content.}
\label{fig:4}
\end{figure}

Figure \ref{fig:4}A reports the partial radial distribution function
$g_{\rm SCN^{-}-SCN^{-}}(r_{\rm CC})$ computed from the MD
trajectories. Here, $r_{\rm CC}$ is the carbon-carbon distance between
the SCN$^-$ anions. The first maximum occurs at $r_{\rm CC} \sim
3.8$~\AA\/ for mixtures with larger water content, especially for 80\%
and larger. This peak does not exist in pure acetamide (blue) and then
increases gradually for mixtures with 20\% to 80\% water content. At
above 50\% water content the peak height increases sharply, by a
factor of close to two. As the vibrational energy transfer rate
depends on the inverse 6th power of the separation of the oscillators,
it is concluded that it is this peak in the $g_{\rm
  SCN^{-}-SCN^{-}}(r_{\rm CC})$ radial distribution function that is
responsible for the cross relaxation. Indeed, the cross-relaxation
rate follows the height of this peak in particular with respect to the
step-like rise beyond 80\% water content (Figure~\ref{fig:2}). It is
also noted that the position of the first maximum in $g_{\rm
  SCN^{-}-SCN^{-}}(r_{\rm CC})$ changes only marginally (except for
the water-free system) but the peak height increases by a factor of
three between 20\% and 95\% water content. This indicates that
predominantly, an angular reorientation of neighboring SCN$^-$ ions
occurs upon increased water content.\\

\noindent
The rise of the first peak in $g_{\rm SCN^{-}-SCN^{-}}(r_{\rm CC})$
correlates with the decrease of the first peaks in $g_{\rm
  SCN^{-}-K^{+}}(r_{\rm CK^{+}})$ at $r_{\rm CK^{+}} = 3.65$~\AA\/ and
4.35~\AA\/ (Figure~\ref{fig:4}B). These peaks are related to K$^+$
interacting with the N- and S-ends in axial position of SCN$^-$,
respectively. The plateau at even shorter distances ($r<3.65$~\AA\/)
originates from K$^+$ in an equatorial position relative to the
SCN$^-$ anion (T-shaped configuration). With increasing water content,
the decrease of the peak amplitudes in Figure~\ref{fig:4}B indicates
preferential solvation of K$^+$ by water molecules.\\

\noindent
Concomitantly, the radial distribution function $g_{\rm
  SCN^{-}-H_2O}(r_{\rm CO})$ for SCN$^{-}$ and the water oxygen atom
in Figure~\ref{fig:4}C indicates preferred interaction of these
molecules. Water solvent and, in particular, the positively charged
H$^{\delta +}$ are more frequently attached along the axial position
to the S atom of SCN$^-$ (second peak in Figure~\ref{fig:4}C) and
preferably to the N atom (first peak in Figure~\ref{fig:4}C), thereby
substituting K$^+$.  The higher the water content, the higher the
proportion of K$^+$ solvated by water molecules which prevents K$^+$
from directly interacting with an SCN$^-$ ion. Equatorial close
attachment of water around SCN$^{-}$ occurs proportionally more often
compared with axial attachment at much higher water ratios (initial
plateau in Figure~\ref{fig:4}C). Conversely, acetamide was found to
attach preferentially in an equatorial position around SCN$^-$ (first
broad peak in Figure~\ref{fig:4}D). Acetamide attached axially by the
partially positively charged amine hydrogen atoms to the N and S atom
of SCN$^-$ becomes prevented by already low water content in the
mixture as observable by the diminishing second and third peak in
Figure~\ref{fig:4}D with increasing water ratio. The rapid increase in
the radial distribution function between SCN$^{-}$ anions at
$3.8$~\AA\/ in Figure~\ref{fig:4}A for water ratios 80\% to 90\%
correlates with a significant change in the system composition. The
ratio between the number of SCN$^-$ anions $N(\rm{SCN}^-)$ and
acetamide molecules $N(\rm{acetamide})$ in the simulation system
changes from $N({\rm SCN}^-) < N({\rm acetamide})$ to $N({\rm SCN}^-)
> N({\rm acetamide})$ for mixtures 80\% and 90\%, respectively.\\

\begin{figure}[t]
\centering
\includegraphics[width=0.99\textwidth]{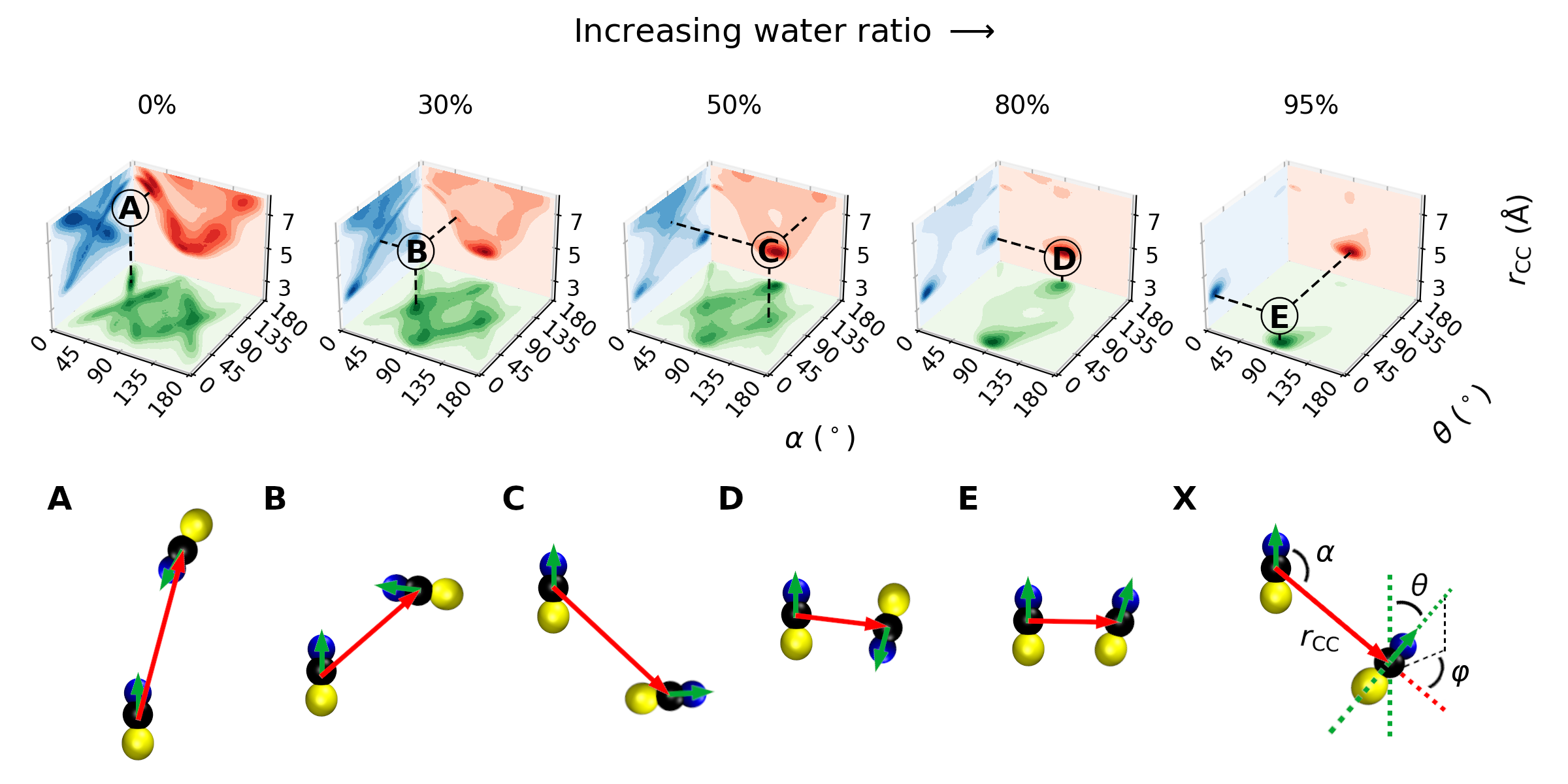}
\caption{SCN$^-$-SCN$^-$ radial-angular alignment distribution SCN$^-$
  anions with increasing water content from left to right. The
  orientation angle $\alpha$ is defined between the CN bond vector
  (green arrow) of the center SCN$^-$ and the C-C connection vector
  (red arrow) to the second SCN$^-$. The alignment $\theta$ denotes
  the angle between both CN bond vectors and $r_\mathrm{CC}$ is the
  distance between the C atoms of neighboring SCN$^-$ anions.  The
  distribution along the $r_\mathrm{CC}$ is normalized as in the
  radial distribution function.  Features A to E correspond to high
  density points in the distribution. The values for $\phi$ in all
  relative orientations $(r, \theta, \alpha)$ are unspecified, see
  panel X.}
\label{fig:5}
\end{figure}

\noindent
The increase of the first maximum of $g_{\rm SCN^{-}-SCN^{-}}(r_{\rm
  CC})$ in Figure~\ref{fig:4}A and shift towards shorter $r_{\rm CC}$
with increasing water content indicates that the arrangement of
SCN$^-$ anions becomes more dense and forms clusters. The denser
packing correlates with increased order in the orientation between
neighboring SCN$^-$ anions. This effect is shown in
Figure~\ref{fig:5}, which reports projections $P(\alpha,\theta)$
(green), $P(\alpha,r_{\rm CC})$ (red), and $P(\theta,r_{\rm CC})$
(blue) of the 3-dimensional probability distribution function
$P(\alpha,\theta,r_{\rm CC})$ of the relative configurations of all
SCN$^-$ anions with carbon-carbon distances $r_{\rm CC} < 8$ \AA\/. In
pure acetamide (left panel Figure~\ref{fig:5}), the angular
orientation between two SCN$^-$ molecules is broadly distributed over
a larger angular range in $\alpha$ and $ \theta$. As the water content
increases (middle and right panels), SCN$^-$ is more inclined for
equatorial stacking with parallel or antiparallel alignment. This
becomes evident as the angular distribution in $\alpha$ (red contour
plot) and $\theta$ (blue contour plot) against the SCN$^-$-SCN$^-$
distance $r_{\rm CC}$ contracts towards a parallel ($\alpha \sim
90^\circ$, $\theta \sim 0^\circ$, Figure~\ref{fig:5}E) and
anti-parallel ($\alpha \sim 90^\circ$, $\theta \sim 180^\circ$,
Figure~\ref{fig:5}D) alignment. Increased order in the angular
orientation for neighboring SCN$^-$ anions increases the tendency for
smaller SCN$^-$-SCN$^-$ separations and leads to the peak at 3.8~\AA\/
in the partial radial distribution function $g_{\rm
  SCN^{-}-SCN^{-}}(r_{\rm CC})$ for increasing water ratios
(Figure~\ref{fig:4}A).\\

\noindent
The transition dipole coupling not only depends on inter-ion distance,
but also on the relative orientation, see Eq.~\ref{eq:BetaCoupling}.
For an equatorial stacking with parallel or antiparallel alignment
(structures D and E in Figure~\ref{fig:5}), the transition dipole
coupling is positive since the first term in
Eq.~\ref{eq:BetaCoupling} dominates given that
$\vec{r}_{ij}\perp\vec{\mu}_{i}=0$. Such structures are called
H-aggregates.\cite{May2004} On the other hand, the second term
overcompensates the first one for a tail-to-tail arrangement
(structure A in Figure~\ref{fig:5}, J-aggregates). In between,
there are magic angle configurations similar to structure B or C,
where both terms cancel each other. Figure~\ref{fig:5} shows that the
most sharply defined structures are in fact H-aggregates with a
population that increases with water content. At the same time, they
reveal the shortest $r_{\rm CC}$ distance, and hence dominate the
cross-relaxation rate. For positive transition dipole coupling, the
higher frequency state will be the symmetric (in-phase) linear
combination of SCN local states, and thus will gain oscillator
strength at the expense of the out-of-phase, lower-frequency
state.\cite{May2004} 2D-IR spectroscopy amplifies this effect due to
the 4th-power dependence of the peak intensity on the transition
dipole (in contrast to a quadratic dependence of linear absorption
spectroscopy). Indeed, the higher frequency band is stronger in the
simulated 2D~IR spectra (Fig.~\ref{fig:1}, bottom) due to this
effect (but one can not rule out that the
KS$^{13}$C$^{15}$N/KS$^{12}$C$^{14}$N mixture has not been exactly
50\%/50\% in the experiment.)\\
 
\noindent
THz spectroscopy is sensitive to structural changes on larger spatial
scales. As an example, {\it ab initio} MD simulations of liquid water
concluded that solutes can impact the hydrogen bond network on length
scales corresponding to at least two layers of solvating water
molecules, i.e. $\sim 10$ \AA, and affect the THz
spectrum.\cite{marx:2010} Here, it is found that the integrated
absorption increases in a nonlinear fashion as water content in the
system increases; see Figure \ref{fig:3}D. Such a nonlinear dependence
indicates that water is not only a dilution component. Rather, the
simulations suggest that as the water content increases,
microheterogeneous patches of water develop. Microheterogeneity has
already been reported for acetonitrile/water mixtures which have been
linked to anomalous thermodynamic properties of such
mixtures.\cite{moreau:1974,reimers:1999,MM.acn:2007} Similarly,
mixtures of very dilute methanol and acetonitrile lead to clustering
of methanol as tetramers or larger clusters\cite{besnard:1992} and
from small angle neutron scattering experiments on sorbitol-water
mixtures which also reported microheterogeneous clustering of the
water.\cite{chou:2012} Here, microheterogeneous clustering of water
molecules can be gleaned from Figure \ref{fig:3}E, which implies that
the integrated THz absorption as a function of the mixing ratio
follows the same nonlinear behaviour as the size distribution of water
clusters in the system. The nonlinearity may arise from the fact that
the number of interactions one specific water molecule in a water
cluster is involved in scales as $\propto n^2$. It is likely that
including polarizability in the simulations - in particular for
water - further improves quantitative agreement with the
experiments. For pure water it is known that point charge models
capture the 600 cm$^{-1}$ absorption correctly whereby the line
shape extends well beyond the 200 cm$^{-1}$ feature due to the
oxygen--oxygen low frequency motion without, however, accounting for
the spectral density due to neglect of polarizability. Including
polarization in the model leads to the spectral response in better
agreement with experiment. However, whether and in which way these
observations for pure water translate to the heterogeneous system
considered in the present work is unclear.\\

\noindent
The $g_{\rm SCN^{-}-SCN^{-}}(r_{\rm CC})$ radial distribution
functions in Figure \ref{fig:4}A show that with increasing hydration
the probability to find two SCN$^{-}$ ions at their most likely
separation ($r_{\rm CC} \sim 3.8$ \AA\/) increases whereas the
position of the maximum is largely unaffected. Based on the
dielectric constants for water (80) compared with acetamide (60) it
would be expected that hydration of SCN$^-$ is facilitated by
increasing water content that results in increased probability to
populate longer $r_{\rm CC}$ separations or a shift of the maximum
value for $P(r_{\rm CC})$ to larger values of $r_{\rm CC}$, or
both. However, for complex and electrostatically demanding mixtures
such as DESs it is likely that expectations governing bulk solvation
do not simply translate. In other words: ``local solvation'' as is
operative in DESs is governed by molecular driving forces which may
result in phenomenologically different observations compared with
what is known from ``bulk solvation''.\\

\section{Conclusion}
In conclusion, spectroscopic responses in two different regions of the
optical spectrum (infrared and THz) for a heterogeneous,
electrostatically dominated system are realistically described from
atomistic simulations for a range of acetamide/water ratios. The
integrated THz absorption increases in a nonlinear fashion due to the
quadratic dependence of the number of interactions in the water
clusters that form with increasing water content and to which this
spectroscopy is sensitive to. Conversely, 2D IR spectroscopy probes
the closer environment of an oscillator (here the CN-stretch) and
coupling between the oscillators. With increasing water content the
relative orientation of two adjacent SCN$^-$ anions changes from
predominantly linear or angled to a more parallel arrangement. The
agreement between simulations and experiments points to an adequate
description of the intra- and intermolecular interactions in the
complex system that can provide insights in the physical and chemical
properties not only for SCN$^-$ clustering in eutectic acetamide:water
solvent but for ion clustering in DESs in general. \\

\begin{acknowledgement}

This work was supported by the University of Basel, the Swiss National
Science Foundation through grants 200021-117810, 200020-188724, the
NCCR MUST, and the European Union's Horizon 2020 research and
innovation program under the Marie Sk{\l}odowska-Curie grant agreement
No 801459 - FP-RESOMUS.\\

\end{acknowledgement}

\begin{suppinfo}

The experimental and computational setup, the sample preparation and  supplementary figures are provided in the supporting information.

\end{suppinfo}

\bibliography{ms}

\providecommand{\latin}[1]{#1}
\makeatletter
\providecommand{\doi}
  {\begingroup\let\do\@makeother\dospecials
  \catcode`\{=1 \catcode`\}=2 \doi@aux}
\providecommand{\doi@aux}[1]{\endgroup\texttt{#1}}
\makeatother
\providecommand*\mcitethebibliography{\thebibliography}
\csname @ifundefined\endcsname{endmcitethebibliography}
  {\let\endmcitethebibliography\endthebibliography}{}
\begin{mcitethebibliography}{95}
\providecommand*\natexlab[1]{#1}
\providecommand*\mciteSetBstSublistMode[1]{}
\providecommand*\mciteSetBstMaxWidthForm[2]{}
\providecommand*\mciteBstWouldAddEndPuncttrue
  {\def\EndOfBibitem{\unskip.}}
\providecommand*\mciteBstWouldAddEndPunctfalse
  {\let\EndOfBibitem\relax}
\providecommand*\mciteSetBstMidEndSepPunct[3]{}
\providecommand*\mciteSetBstSublistLabelBeginEnd[3]{}
\providecommand*\EndOfBibitem{}
\mciteSetBstSublistMode{f}
\mciteSetBstMaxWidthForm{subitem}{(\alph{mcitesubitemcount})}
\mciteSetBstSublistLabelBeginEnd
  {\mcitemaxwidthsubitemform\space}
  {\relax}
  {\relax}

\bibitem[Abbott \latin{et~al.}(2003)Abbott, Capper, Davies, Rasheed, and
  Tambyrajah]{abbott2003DES}
Abbott,~A.~P.; Capper,~G.; Davies,~D.~L.; Rasheed,~R.~K.; Tambyrajah,~V. Novel
  Solvent Properties of Choline Chloride/urea Mixtures. \emph{Chem. Commun.}
  \textbf{2003}, \emph{1}, 70--71\relax
\mciteBstWouldAddEndPuncttrue
\mciteSetBstMidEndSepPunct{\mcitedefaultmidpunct}
{\mcitedefaultendpunct}{\mcitedefaultseppunct}\relax
\EndOfBibitem
\bibitem[Marcus(2019)]{marcus2019trends}
Marcus,~Y. \emph{Deep Eutectic Solvents}; Springer, 2019; pp 185--191\relax
\mciteBstWouldAddEndPuncttrue
\mciteSetBstMidEndSepPunct{\mcitedefaultmidpunct}
{\mcitedefaultendpunct}{\mcitedefaultseppunct}\relax
\EndOfBibitem
\bibitem[Martins \latin{et~al.}(2019)Martins, Pinho, and
  Coutinho]{martins2019defdes}
Martins,~M.~A.; Pinho,~S.~P.; Coutinho,~J.~A. Insights into the Nature of
  Eutectic and Deep Eutectic Mixtures. \emph{J. Solution Chem.} \textbf{2019},
  \emph{48}, 962--982\relax
\mciteBstWouldAddEndPuncttrue
\mciteSetBstMidEndSepPunct{\mcitedefaultmidpunct}
{\mcitedefaultendpunct}{\mcitedefaultseppunct}\relax
\EndOfBibitem
\bibitem[Hansen \latin{et~al.}(2021)Hansen, Spittle, Chen, Poe, Zhang, Klein,
  Horton, Adhikari, Zelovich, Doherty, Gurkan, Maginn, Ragauskas, Dadmun,
  Zawodzinski, Baker, Tuckerman, Savinell, and Sangoro]{hansen:2020}
Hansen,~B.~B. \latin{et~al.}  Deep Eutectic Solvents: A Review of Fundamentals
  and Applications. \emph{Chem. Rev.} \textbf{2021}, \emph{121},
  1232--1285\relax
\mciteBstWouldAddEndPuncttrue
\mciteSetBstMidEndSepPunct{\mcitedefaultmidpunct}
{\mcitedefaultendpunct}{\mcitedefaultseppunct}\relax
\EndOfBibitem
\bibitem[Smith \latin{et~al.}(2014)Smith, Abbott, and Ryder]{Smith2014}
Smith,~E.~L.; Abbott,~A.~P.; Ryder,~K.~S. {Deep Eutectic Solvents (DESs) and
  Their Applications}. \emph{Chem. Rev.} \textbf{2014}, \emph{114},
  11060--11082\relax
\mciteBstWouldAddEndPuncttrue
\mciteSetBstMidEndSepPunct{\mcitedefaultmidpunct}
{\mcitedefaultendpunct}{\mcitedefaultseppunct}\relax
\EndOfBibitem
\bibitem[Lomba \latin{et~al.}(2021)Lomba, Ribate, Sang\"{u}esa, Concha,
  Garralaga, Errazquin, Garc\'{i}a, and Giner]{Lomba2021}
Lomba,~L.; Ribate,~M.{\textsuperscript{a}}~P.; Sang\"{u}esa,~E.; Concha,~J.; Garralaga,~M.
  {\textsuperscript{a}}~P.; Errazquin,~D.; Garc\'{i}a,~C.~B.; Giner,~B. Deep
  Eutectic Solvents: Are They Safe? \emph{App. Sci.} \textbf{2021}, \emph{11},
  10061\relax
\mciteBstWouldAddEndPuncttrue
\mciteSetBstMidEndSepPunct{\mcitedefaultmidpunct}
{\mcitedefaultendpunct}{\mcitedefaultseppunct}\relax
\EndOfBibitem
\bibitem[Emami and Shayanfar(2020)Emami, and Shayanfar]{emami2020desdrugreview}
Emami,~S.; Shayanfar,~A. Deep Eutectic Solvents for Pharmaceutical Formulation
  and Drug Delivery Applications. \emph{Pharm. Dev. Technol.} \textbf{2020},
  \emph{25}, 779--796\relax
\mciteBstWouldAddEndPuncttrue
\mciteSetBstMidEndSepPunct{\mcitedefaultmidpunct}
{\mcitedefaultendpunct}{\mcitedefaultseppunct}\relax
\EndOfBibitem
\bibitem[Mi\v{s}an \latin{et~al.}(2020)Mi\v{s}an, Nadpal, Stupar, Poji\'{c},
  Mandi\'{c}, Verpoorte, and Choi]{misan2019desfoodreview}
Mi\v{s}an,~A.; Nadpal,~J.; Stupar,~A.; Poji\'{c},~M.; Mandi\'{c},~A.;
  Verpoorte,~R.; Choi,~Y.~H. The Perspectives of Natural Deep Eutectic Solvents
  in Agri-Food Sector. \emph{Crit. Rev. Food Sci. Nut.} \textbf{2020},
  \emph{60}, 2564--2592\relax
\mciteBstWouldAddEndPuncttrue
\mciteSetBstMidEndSepPunct{\mcitedefaultmidpunct}
{\mcitedefaultendpunct}{\mcitedefaultseppunct}\relax
\EndOfBibitem
\bibitem[Majid \latin{et~al.}(2020)Majid, Zaid, Kait, Jumbri, Yuan, and
  Rajasuriyan]{majid2020desolireview}
Majid,~M. F.~B.; Zaid,~H. F. B.~M.; Kait,~C.~F.; Jumbri,~K.; Yuan,~L.~C.;
  Rajasuriyan,~S. Futuristic Advance and Perspective of Deep Eutectic Solvent
  for Extractive Desulfurization of Fuel Oil: A Review. \emph{J. Mol. Liq.}
  \textbf{2020}, 112870\relax
\mciteBstWouldAddEndPuncttrue
\mciteSetBstMidEndSepPunct{\mcitedefaultmidpunct}
{\mcitedefaultendpunct}{\mcitedefaultseppunct}\relax
\EndOfBibitem
\bibitem[Cai and Qiu(2019)Cai, and Qiu]{cai2019applicationdesreview}
Cai,~T.; Qiu,~H. Application of Deep Eutectic Solvents in Chromatography: A
  Review. \emph{TrAC, Trends Anal. Chem.} \textbf{2019}, 115623\relax
\mciteBstWouldAddEndPuncttrue
\mciteSetBstMidEndSepPunct{\mcitedefaultmidpunct}
{\mcitedefaultendpunct}{\mcitedefaultseppunct}\relax
\EndOfBibitem
\bibitem[Marcus(2019)]{marcus2019applicationsdesreview}
Marcus,~Y. \emph{Deep Eutectic Solvents}; Springer, 2019; pp 111--151\relax
\mciteBstWouldAddEndPuncttrue
\mciteSetBstMidEndSepPunct{\mcitedefaultmidpunct}
{\mcitedefaultendpunct}{\mcitedefaultseppunct}\relax
\EndOfBibitem
\bibitem[Perna \latin{et~al.}(2019)Perna, Vitale, and
  Capriati]{perna2019desgreenreview}
Perna,~F.~M.; Vitale,~P.; Capriati,~V. Deep Eutectic Solvents and Their
  Applications As Green Solvents. \emph{Curr. Opin. Green Sustain. Chem.}
  \textbf{2019}, \emph{21}, 27--33\relax
\mciteBstWouldAddEndPuncttrue
\mciteSetBstMidEndSepPunct{\mcitedefaultmidpunct}
{\mcitedefaultendpunct}{\mcitedefaultseppunct}\relax
\EndOfBibitem
\bibitem[Amico \latin{et~al.}(1987)Amico, Berchiesi, Cametti, and
  Di~Biasio]{amico1987dielectric}
Amico,~A.; Berchiesi,~G.; Cametti,~C.; Di~Biasio,~A. Dielectric Relaxation
  Spectroscopy of an Acetamide--Sodium Thiocyanate Eutectic Mixture. \emph{J.
  Chem. Soc., Faraday Trans. 2} \textbf{1987}, \emph{83}, 619--626\relax
\mciteBstWouldAddEndPuncttrue
\mciteSetBstMidEndSepPunct{\mcitedefaultmidpunct}
{\mcitedefaultendpunct}{\mcitedefaultseppunct}\relax
\EndOfBibitem
\bibitem[Berchiesi(1999)]{berchiesi1999structural}
Berchiesi,~G. Structural Microheterogeneity and Evidence of Cooperative
  Movement of Charges in Molten Acetamide-electrolyte Mixtures. \emph{J. Mol.
  Liq.} \textbf{1999}, \emph{83}, 271--282\relax
\mciteBstWouldAddEndPuncttrue
\mciteSetBstMidEndSepPunct{\mcitedefaultmidpunct}
{\mcitedefaultendpunct}{\mcitedefaultseppunct}\relax
\EndOfBibitem
\bibitem[Berchiesi \latin{et~al.}(1992)Berchiesi, Farhat, de~Angelis, and
  Barocci]{berchiesi1992high}
Berchiesi,~G.; Farhat,~F.; de~Angelis,~M.; Barocci,~S. High Dielectric Constant
  Supercooled Liquids, Tools in Energetic Problems. \emph{J. Mol. Liq.}
  \textbf{1992}, \emph{54}, 103--113\relax
\mciteBstWouldAddEndPuncttrue
\mciteSetBstMidEndSepPunct{\mcitedefaultmidpunct}
{\mcitedefaultendpunct}{\mcitedefaultseppunct}\relax
\EndOfBibitem
\bibitem[Spittle \latin{et~al.}(2022)Spittle, Poe, Doherty, Kolodziej, Heroux,
  Haque, Squire, Cosby, Zhang, Fraenza, Bhattacharyya, Tyagi, Peng, Elgammal,
  Zawodzinski, Tuckerman, Greenbaum, Gurkan, Burda, Dadmun, Maginn, and
  Sangoro]{spittle:2022}
Spittle,~S. \latin{et~al.}  Evolution of Microscopic Heterogeneity and Dynamics
  in Choline Chloride-based Deep Eutectic Solvents. \emph{Nat. Comm.}
  \textbf{2022}, \emph{13}, 1--14\relax
\mciteBstWouldAddEndPuncttrue
\mciteSetBstMidEndSepPunct{\mcitedefaultmidpunct}
{\mcitedefaultendpunct}{\mcitedefaultseppunct}\relax
\EndOfBibitem
\bibitem[Berchiesi \latin{et~al.}(1995)Berchiesi, Rafaiani, Vitali, and
  Farhat]{berchiesi1995cryoscopic}
Berchiesi,~G.; Rafaiani,~G.; Vitali,~G.; Farhat,~F. Cryoscopic and Dynamic
  Study of the Molten System Fluoroacetamide-sodium Trifluoroacetate. \emph{J.
  Therm. Anal. Calorim.} \textbf{1995}, \emph{44}, 1313--1319\relax
\mciteBstWouldAddEndPuncttrue
\mciteSetBstMidEndSepPunct{\mcitedefaultmidpunct}
{\mcitedefaultendpunct}{\mcitedefaultseppunct}\relax
\EndOfBibitem
\bibitem[Berchiesi \latin{et~al.}(1983)Berchiesi, Vitali, Passamonti, and
  Plowiec]{berchiesi1983viscoelastic}
Berchiesi,~G.; Vitali,~G.; Passamonti,~P.; Plowiec,~R. Viscoelastic Relaxation
  in the Acetamide + Sodium Thiocyanate Binary System. \emph{J. Chem. Soc.,
  Faraday Trans. 2} \textbf{1983}, \emph{79}, 1257--1263\relax
\mciteBstWouldAddEndPuncttrue
\mciteSetBstMidEndSepPunct{\mcitedefaultmidpunct}
{\mcitedefaultendpunct}{\mcitedefaultseppunct}\relax
\EndOfBibitem
\bibitem[Farrat and Berchiesi(1992)Farrat, and Berchiesi]{farrat1992ultrasonic}
Farrat,~F.; Berchiesi,~G. An Ultrasonic Preliminary Study of the Equilibria
  Involved in Solutions of Bipyridil. \emph{J. Mol. Liq.} \textbf{1992},
  \emph{54}, 131--135\relax
\mciteBstWouldAddEndPuncttrue
\mciteSetBstMidEndSepPunct{\mcitedefaultmidpunct}
{\mcitedefaultendpunct}{\mcitedefaultseppunct}\relax
\EndOfBibitem
\bibitem[Biswas \latin{et~al.}(2014)Biswas, Das, and Shirota]{biswas2014rikes}
Biswas,~R.; Das,~A.; Shirota,~H. Low-frequency Collective Dynamics in Deep
  Eutectic Solvents of Acetamide and Electrolytes: A Femtosecond Raman-Induced
  Kerr Effect Spectroscopic Study. \emph{J. Chem. Phys.} \textbf{2014},
  \emph{141}, 134506\relax
\mciteBstWouldAddEndPuncttrue
\mciteSetBstMidEndSepPunct{\mcitedefaultmidpunct}
{\mcitedefaultendpunct}{\mcitedefaultseppunct}\relax
\EndOfBibitem
\bibitem[Gazi \latin{et~al.}(2011)Gazi, Guchhait, Daschakraborty, and
  Biswas]{gazi2011fluorescence}
Gazi,~H. A.~R.; Guchhait,~B.; Daschakraborty,~S.; Biswas,~R. Fluorescence
  Dynamics in Supercooled (Acetamide+ Calcium Nitrate) Molten Mixtures.
  \emph{Chem. Phys. Lett.} \textbf{2011}, \emph{501}, 358--363\relax
\mciteBstWouldAddEndPuncttrue
\mciteSetBstMidEndSepPunct{\mcitedefaultmidpunct}
{\mcitedefaultendpunct}{\mcitedefaultseppunct}\relax
\EndOfBibitem
\bibitem[Guchhait \latin{et~al.}(2010)Guchhait, Al~Rasid~Gazi, Kashyap, and
  Biswas]{guchhait2010fluorescence}
Guchhait,~B.; Al~Rasid~Gazi,~H.; Kashyap,~H.~K.; Biswas,~R. Fluorescence
  Spectroscopic Studies of (acetamide+ Sodium/potassium Thiocyanates) Molten
  Mixtures: Composition and Temperature Dependence. \emph{J. Phys. Chem. B}
  \textbf{2010}, \emph{114}, 5066--5081\relax
\mciteBstWouldAddEndPuncttrue
\mciteSetBstMidEndSepPunct{\mcitedefaultmidpunct}
{\mcitedefaultendpunct}{\mcitedefaultseppunct}\relax
\EndOfBibitem
\bibitem[Guchhait \latin{et~al.}(2012)Guchhait, Daschakraborty, and
  Biswas]{guchhait2012medium}
Guchhait,~B.; Daschakraborty,~S.; Biswas,~R. Medium Decoupling of Dynamics at
  Temperatures~ 100 K above Glass-transition Temperature: A Case Study with
  (acetamide+ Lithium Bromide/nitrate) Melts. \emph{J. Chem. Phys.}
  \textbf{2012}, \emph{136}, 174503\relax
\mciteBstWouldAddEndPuncttrue
\mciteSetBstMidEndSepPunct{\mcitedefaultmidpunct}
{\mcitedefaultendpunct}{\mcitedefaultseppunct}\relax
\EndOfBibitem
\bibitem[Guchhait \latin{et~al.}(2014)Guchhait, Das, Daschakraborty, and
  Biswas]{guchhait2014interaction}
Guchhait,~B.; Das,~S.; Daschakraborty,~S.; Biswas,~R. Interaction and Dynamics
  of (alkylamide+ Electrolyte) Deep Eutectics: Dependence on Alkyl
  Chain-length, Temperature, and Anion Identity. \emph{J. Chem. Phys.}
  \textbf{2014}, \emph{140}, 104514\relax
\mciteBstWouldAddEndPuncttrue
\mciteSetBstMidEndSepPunct{\mcitedefaultmidpunct}
{\mcitedefaultendpunct}{\mcitedefaultseppunct}\relax
\EndOfBibitem
\bibitem[Kumari \latin{et~al.}(2018)Kumari, Shobhna, Kaur, and
  Kashyap]{Kumari2018}
Kumari,~P.; Shobhna,; Kaur,~S.; Kashyap,~H.~K. {Influence of Hydration on the
  Structure of Reline Deep Eutectic Solvent: A Molecular Dynamics Study}.
  \emph{ACS Omega} \textbf{2018}, \emph{3}, 15246--15255\relax
\mciteBstWouldAddEndPuncttrue
\mciteSetBstMidEndSepPunct{\mcitedefaultmidpunct}
{\mcitedefaultendpunct}{\mcitedefaultseppunct}\relax
\EndOfBibitem
\bibitem[Kaur \latin{et~al.}(2019)Kaur, Malik, and Kashyap]{kaur:2019}
Kaur,~S.; Malik,~A.; Kashyap,~H.~K. Anatomy of microscopic structure of
  ethaline deep eutectic solvent decoded through molecular dynamics
  simulations. \emph{J. Phys. Chem. B} \textbf{2019}, \emph{123},
  8291--8299\relax
\mciteBstWouldAddEndPuncttrue
\mciteSetBstMidEndSepPunct{\mcitedefaultmidpunct}
{\mcitedefaultendpunct}{\mcitedefaultseppunct}\relax
\EndOfBibitem
\bibitem[Kaur \latin{et~al.}(2020)Kaur, Gupta, and Kashyap]{kaur:2020}
Kaur,~S.; Gupta,~A.; Kashyap,~H.~K. How hydration affects the microscopic
  structural morphology in a deep eutectic solvent. \emph{J. Phys. Chem. B}
  \textbf{2020}, \emph{124}, 2230--2237\relax
\mciteBstWouldAddEndPuncttrue
\mciteSetBstMidEndSepPunct{\mcitedefaultmidpunct}
{\mcitedefaultendpunct}{\mcitedefaultseppunct}\relax
\EndOfBibitem
\bibitem[Hammond \latin{et~al.}(2017)Hammond, Bowron, and Edler]{Hammond2017}
Hammond,~O.~S.; Bowron,~D.~T.; Edler,~K.~J. {The Effect of Water upon Deep
  Eutectic Solvent Nanostructure: An Unusual Transition from Ionic Mixture to
  Aqueous Solution}. \emph{Angew. Chem. Int. Ed.} \textbf{2017}, \emph{56},
  9782--9785\relax
\mciteBstWouldAddEndPuncttrue
\mciteSetBstMidEndSepPunct{\mcitedefaultmidpunct}
{\mcitedefaultendpunct}{\mcitedefaultseppunct}\relax
\EndOfBibitem
\bibitem[Pal and Biswas(2011)Pal, and Biswas]{pal2011heterogeneity}
Pal,~T.; Biswas,~R. Heterogeneity and Viscosity Decoupling in (Acetamide+
  Electrolyte) Molten Mixtures: A Model Simulation Study. \emph{Chem. Phys.
  Lett.} \textbf{2011}, \emph{517}, 180--185\relax
\mciteBstWouldAddEndPuncttrue
\mciteSetBstMidEndSepPunct{\mcitedefaultmidpunct}
{\mcitedefaultendpunct}{\mcitedefaultseppunct}\relax
\EndOfBibitem
\bibitem[Kaur \latin{et~al.}(2016)Kaur, Gupta, and Kashyap]{Kaur2016}
Kaur,~S.; Gupta,~A.; Kashyap,~H.~K. {Nanoscale Spatial Heterogeneity in Deep
  Eutectic Solvents}. \emph{J. Phys. Chem. B} \textbf{2016}, \emph{120},
  6712--6720\relax
\mciteBstWouldAddEndPuncttrue
\mciteSetBstMidEndSepPunct{\mcitedefaultmidpunct}
{\mcitedefaultendpunct}{\mcitedefaultseppunct}\relax
\EndOfBibitem
\bibitem[Dinda \latin{et~al.}(2021)Dinda, Sil, Das, Tarif, and
  Biswas]{dinda:2021}
Dinda,~S.; Sil,~A.; Das,~A.; Tarif,~E.; Biswas,~R. Does Urea Modify
  Microheterogeneous Nature of Ionic Amide Deep Eutectics? Clues from
  Non-Reactive and Reactive Solute-Centered Dynamics. \emph{J. Mol. Spectrosc.}
  \textbf{2021}, 118126\relax
\mciteBstWouldAddEndPuncttrue
\mciteSetBstMidEndSepPunct{\mcitedefaultmidpunct}
{\mcitedefaultendpunct}{\mcitedefaultseppunct}\relax
\EndOfBibitem
\bibitem[Mukherjee \latin{et~al.}(2015)Mukherjee, Das, Choudhury, Barman, and
  Biswas]{Mukherjee2015}
Mukherjee,~K.; Das,~A.; Choudhury,~S.; Barman,~A.; Biswas,~R. Dielectric
  Relaxations of (Acetamide + Electrolyte) Deep Eutectic Solvents in the
  Frequency Window, 0.2 $\le$ v/GHz $\le$ 50: Anion and Cation Dependence.
  \emph{J. Phys. Chem. B} \textbf{2015}, \emph{119}, 8063--8071\relax
\mciteBstWouldAddEndPuncttrue
\mciteSetBstMidEndSepPunct{\mcitedefaultmidpunct}
{\mcitedefaultendpunct}{\mcitedefaultseppunct}\relax
\EndOfBibitem
\bibitem[Laage and Hynes(2006)Laage, and Hynes]{laage:2006}
Laage,~D.; Hynes,~J.~T. A Molecular Jump Mechanism of Water Reorientation.
  \emph{Science} \textbf{2006}, \emph{311}, 832--835\relax
\mciteBstWouldAddEndPuncttrue
\mciteSetBstMidEndSepPunct{\mcitedefaultmidpunct}
{\mcitedefaultendpunct}{\mcitedefaultseppunct}\relax
\EndOfBibitem
\bibitem[Lee \latin{et~al.}(2013)Lee, Carr, G{\"o}llner, Hamm, and
  Meuwly]{MM.cn:2013}
Lee,~M.~W.; Carr,~J.~K.; G{\"o}llner,~M.; Hamm,~P.; Meuwly,~M. 2D IR Spectra of
  Cyanide in Water Investigated by Molecular Dynamics Simulations. \emph{J.
  Chem. Phys.} \textbf{2013}, \emph{139}, 054506\relax
\mciteBstWouldAddEndPuncttrue
\mciteSetBstMidEndSepPunct{\mcitedefaultmidpunct}
{\mcitedefaultendpunct}{\mcitedefaultseppunct}\relax
\EndOfBibitem
\bibitem[Kuo \latin{et~al.}(2007)Kuo, Vorobyev, Chen, and Hochstrasser]{kuo07}
Kuo,~C.-H.; Vorobyev,~D.~Y.; Chen,~J.; Hochstrasser,~R.~M. Correlation of the
  Vibrations of the Aqueous Azide Ion with the O-H Modes of Bound Water
  Molecules. \emph{J. Phys. Chem. B} \textbf{2007}, \emph{111},
  14028--14033\relax
\mciteBstWouldAddEndPuncttrue
\mciteSetBstMidEndSepPunct{\mcitedefaultmidpunct}
{\mcitedefaultendpunct}{\mcitedefaultseppunct}\relax
\EndOfBibitem
\bibitem[Bian \latin{et~al.}(2011)Bian, Wen, Li, Chen, Han, Sun, Song, Zhuang,
  and Zheng]{bian2011ionclustering}
Bian,~H.; Wen,~X.; Li,~J.; Chen,~H.; Han,~S.; Sun,~X.; Song,~J.; Zhuang,~W.;
  Zheng,~J. Ion Clustering in Aqueous Solutions Probed with Vibrational Energy
  Transfer. \emph{Proc. Natl. Acad. Sci.} \textbf{2011}, \emph{108},
  4737--4742\relax
\mciteBstWouldAddEndPuncttrue
\mciteSetBstMidEndSepPunct{\mcitedefaultmidpunct}
{\mcitedefaultendpunct}{\mcitedefaultseppunct}\relax
\EndOfBibitem
\bibitem[Bian \latin{et~al.}(2011)Bian, Chen, Li, Wen, and
  Zheng]{bian2011nonresonant}
Bian,~H.; Chen,~H.; Li,~J.; Wen,~X.; Zheng,~J. Nonresonant and Resonant
  Mode-Specific Intermolecular Vibrational Energy Transfers in Electrolyte
  Aqueous Solutions. \emph{J. Phys. Chem. A} \textbf{2011}, \emph{115},
  11657--11664\relax
\mciteBstWouldAddEndPuncttrue
\mciteSetBstMidEndSepPunct{\mcitedefaultmidpunct}
{\mcitedefaultendpunct}{\mcitedefaultseppunct}\relax
\EndOfBibitem
\bibitem[Bian \latin{et~al.}(2012)Bian, Li, Zhang, Chen, Zhuang, Gao, and
  Zheng]{bian2012ion}
Bian,~H.; Li,~J.; Zhang,~Q.; Chen,~H.; Zhuang,~W.; Gao,~Y.~Q.; Zheng,~J. Ion
  Segregation in Aqueous Solutions. \emph{J. Phys. Chem. B} \textbf{2012},
  \emph{116}, 14426--14432\relax
\mciteBstWouldAddEndPuncttrue
\mciteSetBstMidEndSepPunct{\mcitedefaultmidpunct}
{\mcitedefaultendpunct}{\mcitedefaultseppunct}\relax
\EndOfBibitem
\bibitem[Bian \latin{et~al.}(2013)Bian, Chen, Zhang, Li, Wen, Zhuang, and
  Zheng]{bian2013cation}
Bian,~H.; Chen,~H.; Zhang,~Q.; Li,~J.; Wen,~X.; Zhuang,~W.; Zheng,~J. Cation
  Effects on Rotational Dynamics of Anions and Water Molecules in Alkali
  (Li$^+$, Na$^+$, K$^+$, Cs$^+$) Thiocyanate (SCN$^-$) Aqueous Solutions.
  \emph{J. Phys. Chem. B} \textbf{2013}, \emph{117}, 7972--7984\relax
\mciteBstWouldAddEndPuncttrue
\mciteSetBstMidEndSepPunct{\mcitedefaultmidpunct}
{\mcitedefaultendpunct}{\mcitedefaultseppunct}\relax
\EndOfBibitem
\bibitem[Chen \latin{et~al.}(2012)Chen, Bian, Li, Wen, and
  Zheng]{chen2012ultrafast}
Chen,~H.; Bian,~H.; Li,~J.; Wen,~X.; Zheng,~J. Ultrafast Multiple-Mode
  Multiple-dimensional Vibrational Spectroscopy. \emph{Int. Rev. Phys. Chem.}
  \textbf{2012}, \emph{31}, 469--565\relax
\mciteBstWouldAddEndPuncttrue
\mciteSetBstMidEndSepPunct{\mcitedefaultmidpunct}
{\mcitedefaultendpunct}{\mcitedefaultseppunct}\relax
\EndOfBibitem
\bibitem[Chen \latin{et~al.}(2014)Chen, Wen, Li, and Zheng]{Chen2014b}
Chen,~H.; Wen,~X.; Li,~J.; Zheng,~J. Molecular Distances Determined with
  Resonant Vibrational Energy Transfers. \emph{J. Phys. Chem. A} \textbf{2014},
  \emph{118}, 2463--2469\relax
\mciteBstWouldAddEndPuncttrue
\mciteSetBstMidEndSepPunct{\mcitedefaultmidpunct}
{\mcitedefaultendpunct}{\mcitedefaultseppunct}\relax
\EndOfBibitem
\bibitem[Chen \latin{et~al.}(2014)Chen, Wen, Guo, and Zheng]{Chen2014a}
Chen,~H.; Wen,~X.; Guo,~X.; Zheng,~J. {Intermolecular vibrational energy
  transfers in liquids and solids}. \emph{Phys. Chem. Chem. Phys.}
  \textbf{2014}, \emph{16}, 13995--14014\relax
\mciteBstWouldAddEndPuncttrue
\mciteSetBstMidEndSepPunct{\mcitedefaultmidpunct}
{\mcitedefaultendpunct}{\mcitedefaultseppunct}\relax
\EndOfBibitem
\bibitem[Fern{\'{a}}ndez-Ter{\'{a}}n and Hamm(2020)Fern{\'{a}}ndez-Ter{\'{a}}n,
  and Hamm]{Fernandez-Teran2020a}
Fern{\'{a}}ndez-Ter{\'{a}}n,~R.; Hamm,~P. {A Closer Look Into the Distance
  Dependence of Vibrational Energy Transfer on Surfaces Using 2D ATR-IR
  Spectroscopy}. \emph{J. Chem. Phys.} \textbf{2020}, \emph{153}, 154706\relax
\mciteBstWouldAddEndPuncttrue
\mciteSetBstMidEndSepPunct{\mcitedefaultmidpunct}
{\mcitedefaultendpunct}{\mcitedefaultseppunct}\relax
\EndOfBibitem
\bibitem[Schirmer \latin{et~al.}(1970)Schirmer, Noggle, Davis, and
  Hart]{Schirmer1970}
Schirmer,~R.~E.; Noggle,~J.~H.; Davis,~J.~P.; Hart,~P.~A. Determination of
  Molecular Geometry by Quantitative Application of the Nuclear Overhauser
  Effect. \emph{J. Am. Chem. Soc.} \textbf{1970}, \emph{92}, 3266--3273\relax
\mciteBstWouldAddEndPuncttrue
\mciteSetBstMidEndSepPunct{\mcitedefaultmidpunct}
{\mcitedefaultendpunct}{\mcitedefaultseppunct}\relax
\EndOfBibitem
\bibitem[Bell and Saunders(1970)Bell, and Saunders]{Bell1970}
Bell,~R.~A.; Saunders,~J.~K. Correlation of the Intramolecular Nuclear
  Overhauser Effect with Internuclear Distance. \emph{Can. J. Chem.}
  \textbf{1970}, \emph{48}, 1114--1122\relax
\mciteBstWouldAddEndPuncttrue
\mciteSetBstMidEndSepPunct{\mcitedefaultmidpunct}
{\mcitedefaultendpunct}{\mcitedefaultseppunct}\relax
\EndOfBibitem
\bibitem[Stryer and Haugland(1967)Stryer, and Haugland]{Stryer1967}
Stryer,~L.; Haugland,~R.~P. Energy Transfer: A Spectroscopic Ruler. \emph{Proc.
  Natl. Acad. Sci. U. S. A.} \textbf{1967}, \emph{58}, 719--726\relax
\mciteBstWouldAddEndPuncttrue
\mciteSetBstMidEndSepPunct{\mcitedefaultmidpunct}
{\mcitedefaultendpunct}{\mcitedefaultseppunct}\relax
\EndOfBibitem
\bibitem[Cunsolo(2015)]{Cunsolo2015}
Cunsolo,~A. The THz Spectrum of Density Fluctuations of Water: The Viscoelastic
  Regime. \emph{Adv. Condens. Matter Phys.} \textbf{2015}, \emph{2015},
  137435\relax
\mciteBstWouldAddEndPuncttrue
\mciteSetBstMidEndSepPunct{\mcitedefaultmidpunct}
{\mcitedefaultendpunct}{\mcitedefaultseppunct}\relax
\EndOfBibitem
\bibitem[Smolyanskaya \latin{et~al.}(2018)Smolyanskaya, Chernomyrdin, Konovko,
  Zaytsev, Ozheredov, Cherkasova, Nazarov, Guillet, Kozlov, Kistenev, Coutaz,
  Mounaix, Vaks, Son, Cheon, Wallace, Feldman, Popov, Yaroslavsky, Shkurinov,
  and Tuchin]{smolyanskaya2018}
Smolyanskaya,~O. \latin{et~al.}  Terahertz Biophotonics As a Tool for Studies
  of Dielectric and Spectral Properties of Biological Tissues and Liquids.
  \emph{Prog. Quantum Electron.} \textbf{2018}, \emph{62}, 1--77\relax
\mciteBstWouldAddEndPuncttrue
\mciteSetBstMidEndSepPunct{\mcitedefaultmidpunct}
{\mcitedefaultendpunct}{\mcitedefaultseppunct}\relax
\EndOfBibitem
\bibitem[Schotte \latin{et~al.}(2003)Schotte, Lim, Jackson, Smirnov, Soman,
  Olson, Phillips, Wulff, and Anfinrud]{anfinrud:2003}
Schotte,~F.; Lim,~M.; Jackson,~T.~A.; Smirnov,~A.~V.; Soman,~J.; Olson,~J.~S.;
  Phillips,~G.~N.; Wulff,~M.; Anfinrud,~P.~A. Watching a Protein As It
  Functions with 150-ps Time-resolved X-ray Crystallography. \emph{Science}
  \textbf{2003}, \emph{300}, 1944--1947\relax
\mciteBstWouldAddEndPuncttrue
\mciteSetBstMidEndSepPunct{\mcitedefaultmidpunct}
{\mcitedefaultendpunct}{\mcitedefaultseppunct}\relax
\EndOfBibitem
\bibitem[Nutt and Meuwly(2004)Nutt, and Meuwly]{MM.mbco:2004}
Nutt,~D.~R.; Meuwly,~M. Co Migration in Native and Mutant Myoglobin: Atomistic
  Simulations for the Understanding of Protein Function. \emph{Proc. Natl.
  Acad. Sci.} \textbf{2004}, \emph{101}, 5998--6002\relax
\mciteBstWouldAddEndPuncttrue
\mciteSetBstMidEndSepPunct{\mcitedefaultmidpunct}
{\mcitedefaultendpunct}{\mcitedefaultseppunct}\relax
\EndOfBibitem
\bibitem[Castner~Jr and Wishart(2010)Castner~Jr, and Wishart]{castner:2010}
Castner~Jr,~E.~W.; Wishart,~J.~F. Spotlight on Ionic Liquids. \emph{J. Chem.
  Phys.} \textbf{2010}, \emph{132}, 120901\relax
\mciteBstWouldAddEndPuncttrue
\mciteSetBstMidEndSepPunct{\mcitedefaultmidpunct}
{\mcitedefaultendpunct}{\mcitedefaultseppunct}\relax
\EndOfBibitem
\bibitem[Castner~Jr \latin{et~al.}(2011)Castner~Jr, Margulis, Maroncelli, and
  Wishart]{castner:2011}
Castner~Jr,~E.~W.; Margulis,~C.~J.; Maroncelli,~M.; Wishart,~J.~F. Ionic
  Liquids: Structure and Photochemical Reactions. \emph{Ann. Rev. Phys. Chem.}
  \textbf{2011}, \emph{62}, 85--105\relax
\mciteBstWouldAddEndPuncttrue
\mciteSetBstMidEndSepPunct{\mcitedefaultmidpunct}
{\mcitedefaultendpunct}{\mcitedefaultseppunct}\relax
\EndOfBibitem
\bibitem[Kalita \latin{et~al.}(1998)Kalita, Rohman, and Mahiuddin]{Kalita1998}
Kalita,~G.; Rohman,~N.; Mahiuddin,~S. {Viscosity and Molar Volume of Potassium
  Thiocyanate + Sodium Thiocyanate + Acetamide Melt Systems}. \emph{J. Chem.
  Eng.Data} \textbf{1998}, \emph{43}, 148--151\relax
\mciteBstWouldAddEndPuncttrue
\mciteSetBstMidEndSepPunct{\mcitedefaultmidpunct}
{\mcitedefaultendpunct}{\mcitedefaultseppunct}\relax
\EndOfBibitem
\bibitem[Isaac and Kerridge(1988)Isaac, and Kerridge]{DT9880002701}
Isaac,~I.~Y.; Kerridge,~D.~H. {Molten Acetamide-Potassium Thiocyanate Eutectic:
  Spectroscopy of First-row Transition Metal Compounds in a Room Temperature
  Melt}. \emph{J. Chem. Soc.{,} Dalton Trans.} \textbf{1988}, 2701--2704\relax
\mciteBstWouldAddEndPuncttrue
\mciteSetBstMidEndSepPunct{\mcitedefaultmidpunct}
{\mcitedefaultendpunct}{\mcitedefaultseppunct}\relax
\EndOfBibitem
\bibitem[Hu \latin{et~al.}(2004)Hu, Li, Huang, and Chen]{HU200428}
Hu,~Y.; Li,~H.; Huang,~X.; Chen,~L. {Novel Room Temperature Molten Salt
  Electrolyte Based on LiTFSI and Acetamide for Lithium Batteries}.
  \emph{Electrochem. Comm.} \textbf{2004}, \emph{6}, 28--32\relax
\mciteBstWouldAddEndPuncttrue
\mciteSetBstMidEndSepPunct{\mcitedefaultmidpunct}
{\mcitedefaultendpunct}{\mcitedefaultseppunct}\relax
\EndOfBibitem
\bibitem[Wallace(1972)]{Wallace1972}
Wallace,~R.~A. {Solubility of Potassium Halides in Fused Acetamide}.
  \emph{Inorg. Chem.} \textbf{1972}, \emph{11}, 414--415\relax
\mciteBstWouldAddEndPuncttrue
\mciteSetBstMidEndSepPunct{\mcitedefaultmidpunct}
{\mcitedefaultendpunct}{\mcitedefaultseppunct}\relax
\EndOfBibitem
\bibitem[Palmelund \latin{et~al.}(2021)Palmelund, Rantanen, and
  L{\"{o}}bmann]{Palmelund2021}
Palmelund,~H.; Rantanen,~J.; L{\"{o}}bmann,~K. Deliquescence Behavior of Deep
  Eutectic Solvents. \emph{App. Sci.} \textbf{2021}, \emph{11}, 1601\relax
\mciteBstWouldAddEndPuncttrue
\mciteSetBstMidEndSepPunct{\mcitedefaultmidpunct}
{\mcitedefaultendpunct}{\mcitedefaultseppunct}\relax
\EndOfBibitem
\bibitem[Dai \latin{et~al.}(2015)Dai, Witkamp, Verpoorte, and Choi]{DAI2015}
Dai,~Y.; Witkamp,~G.-J.; Verpoorte,~R.; Choi,~Y.~H. {Tailoring Properties of
  Natural Deep Eutectic Solvents with Water to Facilitate their Applications}.
  \emph{Food Chem.} \textbf{2015}, \emph{187}, 14--19\relax
\mciteBstWouldAddEndPuncttrue
\mciteSetBstMidEndSepPunct{\mcitedefaultmidpunct}
{\mcitedefaultendpunct}{\mcitedefaultseppunct}\relax
\EndOfBibitem
\bibitem[Sakpal \latin{et~al.}(2021)Sakpal, Deshmukh, Chatterjee, Ghosh, and
  Bagchi]{bagchi:2021}
Sakpal,~S.~S.; Deshmukh,~S.~H.; Chatterjee,~S.; Ghosh,~D.; Bagchi,~S.
  Transition of a Deep Eutectic Solution to Aqueous Solution: A Dynamical
  Perspective of the Dissolved Solute. \emph{J. Phys. Chem. Lett.}
  \textbf{2021}, \emph{12}, 8784--8789\relax
\mciteBstWouldAddEndPuncttrue
\mciteSetBstMidEndSepPunct{\mcitedefaultmidpunct}
{\mcitedefaultendpunct}{\mcitedefaultseppunct}\relax
\EndOfBibitem
\bibitem[Helbing and Hamm(2011)Helbing, and Hamm]{Helbing2011}
Helbing,~J.; Hamm,~P. Compact Implementation of Fourier Transform
  Two-Dimensional IR Spectroscopy without Phase Ambiguity. \emph{J. Opt. Soc.
  Am. B} \textbf{2011}, \emph{28}, 171--178\relax
\mciteBstWouldAddEndPuncttrue
\mciteSetBstMidEndSepPunct{\mcitedefaultmidpunct}
{\mcitedefaultendpunct}{\mcitedefaultseppunct}\relax
\EndOfBibitem
\bibitem[Hamm \latin{et~al.}(2000)Hamm, Kaindl, and Stenger]{hamm2000noise}
Hamm,~P.; Kaindl,~R.~A.; Stenger,~J. Noise Suppression in Femtosecond
  Mid-infrared Light Sources. \emph{Opt. Lett.} \textbf{2000}, \emph{25},
  1798--1800\relax
\mciteBstWouldAddEndPuncttrue
\mciteSetBstMidEndSepPunct{\mcitedefaultmidpunct}
{\mcitedefaultendpunct}{\mcitedefaultseppunct}\relax
\EndOfBibitem
\bibitem[Bloem \latin{et~al.}(2010)Bloem, Garrett-Roe, Strzalka, Hamm, and
  Donaldson]{bloem10}
Bloem,~R.; Garrett-Roe,~S.; Strzalka,~H.; Hamm,~P.; Donaldson,~P. Enhancing
  Signal Detection and Completely Eliminating Scattering Using
  Quasi-Phase-Cycling in 2D IR Experiments. \emph{Opt. Express} \textbf{2010},
  \emph{18}, 27067--27078\relax
\mciteBstWouldAddEndPuncttrue
\mciteSetBstMidEndSepPunct{\mcitedefaultmidpunct}
{\mcitedefaultendpunct}{\mcitedefaultseppunct}\relax
\EndOfBibitem
\bibitem[Brooks \latin{et~al.}(2009)Brooks, Brooks~III, MacKerell~Jr., Nilsson,
  Petrella, Roux, Won, Archontis, Bartels, Boresch, Caflisch, Caves, Cui,
  Dinner, Feig, Fischer, Gao, Hodoscek, Im, Kuczera, Lazaridis, Ma,
  Ovchinnikov, Paci, Pastor, Post, Schaefer, Tidor, Venable, Woodcock, Wu,
  Yang, York, and Karplus]{Charmm-Brooks-2009}
Brooks,~B.~R. \latin{et~al.}  Charmm: The Biomolecular Simulation Program.
  \emph{J. Comp. Chem.} \textbf{2009}, \emph{30}, 1545--1614\relax
\mciteBstWouldAddEndPuncttrue
\mciteSetBstMidEndSepPunct{\mcitedefaultmidpunct}
{\mcitedefaultendpunct}{\mcitedefaultseppunct}\relax
\EndOfBibitem
\bibitem[Guvench \latin{et~al.}(2011)Guvench, Mallajosyula, Raman, Hatcher,
  Vanommeslaeghe, Foster, Jamison, and MacKerell]{CHARMMFF-ALL36-Guvench2011}
Guvench,~O.; Mallajosyula,~S.~S.; Raman,~E.~P.; Hatcher,~E.;
  Vanommeslaeghe,~K.; Foster,~T.~J.; Jamison,~F.~W.,~II; MacKerell,~A.~D.,~Jr.
  Charmm Additive All-Atom Force Field for Carbohydrate Derivatives and Its
  Utility in Polysaccharide and Carbohydrate-Protein Modeling. \emph{J. Chem.
  Theo. Comp.} \textbf{2011}, \emph{7}, 3162--3180\relax
\mciteBstWouldAddEndPuncttrue
\mciteSetBstMidEndSepPunct{\mcitedefaultmidpunct}
{\mcitedefaultendpunct}{\mcitedefaultseppunct}\relax
\EndOfBibitem
\bibitem[Vanommeslaeghe \latin{et~al.}(2010)Vanommeslaeghe, Hatcher, Acharya,
  Kundu, Zhong, Shim, Darian, Guvench, Lopes, Vorobyov, and
  Mackerell~Jr.]{CGenFF}
Vanommeslaeghe,~K.; Hatcher,~E.; Acharya,~C.; Kundu,~S.; Zhong,~S.; Shim,~J.;
  Darian,~E.; Guvench,~O.; Lopes,~P.; Vorobyov,~I.; Mackerell~Jr.,~A.~D. Charmm
  General Force Field: A Force Field for Drug-Like Molecules Compatible with
  the Charmm All-Atom Additive Biological Force Fields. \emph{J. Comp. Chem.}
  \textbf{2010}, \emph{31}, 671--690\relax
\mciteBstWouldAddEndPuncttrue
\mciteSetBstMidEndSepPunct{\mcitedefaultmidpunct}
{\mcitedefaultendpunct}{\mcitedefaultseppunct}\relax
\EndOfBibitem
\bibitem[Jorgensen \latin{et~al.}(1983)Jorgensen, Chandrasekhar, Madura, Impey,
  and Klein]{TIP3P-Jorgensen-1983}
Jorgensen,~W.~L.; Chandrasekhar,~J.; Madura,~J.~D.; Impey,~R.~W.; Klein,~M.~L.
  Comparison of Simple Potential Functions for Simulating Liquid Water.
  \emph{J. Chem. Phys.} \textbf{1983}, \emph{79}, 926--935\relax
\mciteBstWouldAddEndPuncttrue
\mciteSetBstMidEndSepPunct{\mcitedefaultmidpunct}
{\mcitedefaultendpunct}{\mcitedefaultseppunct}\relax
\EndOfBibitem
\bibitem[Bereau \latin{et~al.}(2013)Bereau, Kramer, and Meuwly]{MM.mtp:2013}
Bereau,~T.; Kramer,~C.; Meuwly,~M. Leveraging Symmetries of Static Atomic
  Multipole Electrostatics in Molecular Dynamics Simulations. \emph{J. Chem.
  Theo. Comput.} \textbf{2013}, \emph{9}, 5450--5459\relax
\mciteBstWouldAddEndPuncttrue
\mciteSetBstMidEndSepPunct{\mcitedefaultmidpunct}
{\mcitedefaultendpunct}{\mcitedefaultseppunct}\relax
\EndOfBibitem
\bibitem[Hedin \latin{et~al.}(2016)Hedin, El~Hage, and Meuwly]{mm.mtp2:2016}
Hedin,~F.; El~Hage,~K.; Meuwly,~M. A Toolkit to Fit Nonbonded Parameters from
  and for Condensed Phase Simulations. \emph{J. Chem. Theo. Comp.}
  \textbf{2016}, \emph{56}, 1479--1489\relax
\mciteBstWouldAddEndPuncttrue
\mciteSetBstMidEndSepPunct{\mcitedefaultmidpunct}
{\mcitedefaultendpunct}{\mcitedefaultseppunct}\relax
\EndOfBibitem
\bibitem[Hedin \latin{et~al.}(2017)Hedin, El~Hage, and Meuwly]{mm.mtp:2017}
Hedin,~F.; El~Hage,~K.; Meuwly,~M. A Toolkit to Fit Nonbonded Parameters from
  and for Condensed Phase Simulations. \emph{J. Chem. Theo. Comp.}
  \textbf{2017}, \emph{57}, 102--103\relax
\mciteBstWouldAddEndPuncttrue
\mciteSetBstMidEndSepPunct{\mcitedefaultmidpunct}
{\mcitedefaultendpunct}{\mcitedefaultseppunct}\relax
\EndOfBibitem
\bibitem[Frisch \latin{et~al.}()Frisch, Trucks, Schlegel, Scuseria, Robb,
  Cheeseman, Scalmani, Barone, Mennucci, Petersson, Nakatsuji, Caricato, Li,
  Hratchian, Izmaylov, Bloino, Zheng, Sonnenberg, Hada, Ehara, Toyota, Fukuda,
  Hasegawa, Ishida, Nakajima, Honda, Kitao, Nakai, Vreven, Montgomery, Jr.,
  Peralta, Ogliaro, Bearpark, Heyd, Brothers, Kudin, Staroverov, Kobayashi,
  Normand, Raghavachari, Rendell, Burant, Iyengar, Tomasi, Cossi, Rega, Millam,
  Klene, Knox, Cross, Bakken, Adamo, Jaramillo, Gomperts, Stratmann, Yazyev,
  Austin, Cammi, Pomelli, Ochterski, Martin, Morokuma, Zakrzewski, Voth,
  Salvador, Dannenberg, Dapprich, Daniels, Farkas, Foresman, Ortiz, Cioslowski,
  , and Fox]{gaussian09}
Frisch,~M.~J. \latin{et~al.}  Gaussian 09, \uppercase{R}evision
  \uppercase{A}.02. \uppercase{G}aussian, Inc., Wallingford, CT, 2009\relax
\mciteBstWouldAddEndPuncttrue
\mciteSetBstMidEndSepPunct{\mcitedefaultmidpunct}
{\mcitedefaultendpunct}{\mcitedefaultseppunct}\relax
\EndOfBibitem
\bibitem[Ryckaert \latin{et~al.}(1977)Ryckaert, Ciccotti, and
  Berendsen]{shake77}
Ryckaert,~J.-P.; Ciccotti,~G.; Berendsen,~H. J.~C. Numerical Integration of the
  Cartesian Equations of Motion of a System with Constraints: Molecular
  Dynamics of N-Alkanes. \emph{J. Chem. Phys.} \textbf{1977}, \emph{23},
  327--341\relax
\mciteBstWouldAddEndPuncttrue
\mciteSetBstMidEndSepPunct{\mcitedefaultmidpunct}
{\mcitedefaultendpunct}{\mcitedefaultseppunct}\relax
\EndOfBibitem
\bibitem[Darden \latin{et~al.}(1993)Darden, York, and Pedersen]{Darden1993}
Darden,~T.; York,~D.; Pedersen,~L. Particle Mesh Ewald: An N$\cdot$log(n)
  Method for Ewald Sums in Large Systems. \emph{J. Chem. Phys.} \textbf{1993},
  \emph{98}, 10089--10092\relax
\mciteBstWouldAddEndPuncttrue
\mciteSetBstMidEndSepPunct{\mcitedefaultmidpunct}
{\mcitedefaultendpunct}{\mcitedefaultseppunct}\relax
\EndOfBibitem
\bibitem[Mart\'{i}nez \latin{et~al.}(2009)Mart\'{i}nez, Andrade, Birgin, and
  Mart\'{i}nez]{martinez:2009}
Mart\'{i}nez,~L.; Andrade,~R.; Birgin,~E.~G.; Mart\'{i}nez,~J.~M. Packmol: A
  Package for Building Initial Configurations for Molecular Dynamics
  Simulations. \emph{J. Comp. Chem.} \textbf{2009}, \emph{30}, 2157--2164\relax
\mciteBstWouldAddEndPuncttrue
\mciteSetBstMidEndSepPunct{\mcitedefaultmidpunct}
{\mcitedefaultendpunct}{\mcitedefaultseppunct}\relax
\EndOfBibitem
\bibitem[Hoover(1985)]{Hoover1985}
Hoover,~W.~G. Canonical Dynamics: Equilibrium Phase-Space Distributions.
  \emph{Phys. Rev. A} \textbf{1985}, \emph{31}, 1695--1697\relax
\mciteBstWouldAddEndPuncttrue
\mciteSetBstMidEndSepPunct{\mcitedefaultmidpunct}
{\mcitedefaultendpunct}{\mcitedefaultseppunct}\relax
\EndOfBibitem
\bibitem[Cho \latin{et~al.}(1994)Cho, Fleming, Saito, Ohmine, and
  Stratt]{stratt:1994}
Cho,~M.; Fleming,~G.~R.; Saito,~S.; Ohmine,~I.; Stratt,~R.~M. Instantaneous
  Normal Mode Analysis of Liquid Water. \emph{J. Chem. Phys.} \textbf{1994},
  \emph{100}, 6672--6683\relax
\mciteBstWouldAddEndPuncttrue
\mciteSetBstMidEndSepPunct{\mcitedefaultmidpunct}
{\mcitedefaultendpunct}{\mcitedefaultseppunct}\relax
\EndOfBibitem
\bibitem[Keyes(1997)]{keyes:1997}
Keyes,~T. Instantaneous Normal Mode Approach to Liquid State Dynamics. \emph{J.
  Phys. Chem. A} \textbf{1997}, \emph{101}, 2921--2930\relax
\mciteBstWouldAddEndPuncttrue
\mciteSetBstMidEndSepPunct{\mcitedefaultmidpunct}
{\mcitedefaultendpunct}{\mcitedefaultseppunct}\relax
\EndOfBibitem
\bibitem[Kindt and Schmuttenmaer(1997)Kindt, and
  Schmuttenmaer]{schmuttenmaer:1997}
Kindt,~J.~T.; Schmuttenmaer,~C.~A. Far-infrared Absorption Spectra of Wwater,
  Ammonia, and Chloroform Calculated from Instantaneous Normal Mode Theory.
  \emph{J. Chem. Phys.} \textbf{1997}, \emph{106}, 4389--4400\relax
\mciteBstWouldAddEndPuncttrue
\mciteSetBstMidEndSepPunct{\mcitedefaultmidpunct}
{\mcitedefaultendpunct}{\mcitedefaultseppunct}\relax
\EndOfBibitem
\bibitem[Imoto \latin{et~al.}(2013)Imoto, Xantheas, and Saito]{saito:2013}
Imoto,~S.; Xantheas,~S.~S.; Saito,~S. Molecular Origin of the Difference in the
  HOH Bend of the IR Spectra between Liquid Water and Ice. \emph{J. Chem.
  Phys.} \textbf{2013}, \emph{138}, 054506\relax
\mciteBstWouldAddEndPuncttrue
\mciteSetBstMidEndSepPunct{\mcitedefaultmidpunct}
{\mcitedefaultendpunct}{\mcitedefaultseppunct}\relax
\EndOfBibitem
\bibitem[Salehi \latin{et~al.}(2019)Salehi, Koner, and Meuwly]{MM.n3:2019}
Salehi,~S.~M.; Koner,~D.; Meuwly,~M. Vibrational Spectroscopy of N$_3^-$ in the
  Gas and Condensed Phase. \emph{J. Phys. Chem. B} \textbf{2019}, \emph{123},
  3282--3290\relax
\mciteBstWouldAddEndPuncttrue
\mciteSetBstMidEndSepPunct{\mcitedefaultmidpunct}
{\mcitedefaultendpunct}{\mcitedefaultseppunct}\relax
\EndOfBibitem
\bibitem[Hamm and Zanni(2011)Hamm, and Zanni]{hamm11}
Hamm,~P.; Zanni,~M.~T. \emph{Concepts and Methods of 2D Infrared Spectroscopy};
  Cambridge University Press: Cambridge, 2011\relax
\mciteBstWouldAddEndPuncttrue
\mciteSetBstMidEndSepPunct{\mcitedefaultmidpunct}
{\mcitedefaultendpunct}{\mcitedefaultseppunct}\relax
\EndOfBibitem
\bibitem[Liang and Jansen(2012)Liang, and Jansen]{Liang2012}
Liang,~C.; Jansen,~T.~L. An Efficient N3-Scaling Propagation Scheme for
  Simulating Two-Dimensional Infrared and Visible Spectra. \emph{J. Chem.
  Theory Comput.} \textbf{2012}, \emph{8}, 1706--1713\relax
\mciteBstWouldAddEndPuncttrue
\mciteSetBstMidEndSepPunct{\mcitedefaultmidpunct}
{\mcitedefaultendpunct}{\mcitedefaultseppunct}\relax
\EndOfBibitem
\bibitem[Tal-Ezer and Kosloff(1984)Tal-Ezer, and Kosloff]{Tal-Ezer1984}
Tal-Ezer,~H.; Kosloff,~R. An Accurate and Efficient Scheme for Propagating the
  Time Dependent Schr{\"{o}}dinger Equation. \emph{J. Chem. Phys.}
  \textbf{1984}, \emph{81}, 3967--3971\relax
\mciteBstWouldAddEndPuncttrue
\mciteSetBstMidEndSepPunct{\mcitedefaultmidpunct}
{\mcitedefaultendpunct}{\mcitedefaultseppunct}\relax
\EndOfBibitem
\bibitem[Guo \latin{et~al.}(2015)Guo, Pagano, Li, Kohen, and Cheatum]{Guo2015}
Guo,~Q.; Pagano,~P.; Li,~Y.~L.; Kohen,~A.; Cheatum,~C.~M. Line Shape Analysis
  of Two-Dimensional Infrared Spectra. \emph{J. Chem. Phys.} \textbf{2015},
  \emph{142}, 212427\relax
\mciteBstWouldAddEndPuncttrue
\mciteSetBstMidEndSepPunct{\mcitedefaultmidpunct}
{\mcitedefaultendpunct}{\mcitedefaultseppunct}\relax
\EndOfBibitem
\bibitem[Kwak \latin{et~al.}(2006)Kwak, Zheng, Cang, and Fayer]{Kwak2006}
Kwak,~K.; Zheng,~J.; Cang,~H.; Fayer,~M.~D. Ultrafast Two-Dimensional Infrared
  Vibrational Echo Chemical Exchange Experiments and Theory. \emph{J. Phys.
  Chem. B} \textbf{2006}, \emph{110}, 19998--20013\relax
\mciteBstWouldAddEndPuncttrue
\mciteSetBstMidEndSepPunct{\mcitedefaultmidpunct}
{\mcitedefaultendpunct}{\mcitedefaultseppunct}\relax
\EndOfBibitem
\bibitem[Ram{\i}rez \latin{et~al.}(2004)Ram{\i}rez, L{\'o}pez-Ciudad, Kumar~P,
  and Marx]{ramrez04}
Ram{\i}rez,~R.; L{\'o}pez-Ciudad,~T.; Kumar~P,~P.; Marx,~D. Quantum Corrections
  to Classical Time-Correlation Functions: Hydrogen Bonding and Anharmonic
  Floppy Modes. \emph{J. Chem. Phys.} \textbf{2004}, \emph{121},
  3973--3983\relax
\mciteBstWouldAddEndPuncttrue
\mciteSetBstMidEndSepPunct{\mcitedefaultmidpunct}
{\mcitedefaultendpunct}{\mcitedefaultseppunct}\relax
\EndOfBibitem
\bibitem[Liu and Liu(2016)Liu, and Liu]{liu2016properties}
Liu,~B.; Liu,~Y. Properties for Binary Mixtures of (Acetamide+ KSCN) Eutectic
  Ionic Liquid with Ethanol at Several Temperatures. \emph{J. Chem. Thermodyn.}
  \textbf{2016}, \emph{92}, 1--7\relax
\mciteBstWouldAddEndPuncttrue
\mciteSetBstMidEndSepPunct{\mcitedefaultmidpunct}
{\mcitedefaultendpunct}{\mcitedefaultseppunct}\relax
\EndOfBibitem
\bibitem[Liu \latin{et~al.}(2013)Liu, Zhao, and Wei]{liu:2013}
Liu,~B.; Zhao,~J.; Wei,~F. Effects of Water on the Properties of Acetamide-KSCN
  Eutectic Ionic Liquids at Several Temperatures. \emph{J. Mol. Liq.}
  \textbf{2013}, \emph{187}, 309--313\relax
\mciteBstWouldAddEndPuncttrue
\mciteSetBstMidEndSepPunct{\mcitedefaultmidpunct}
{\mcitedefaultendpunct}{\mcitedefaultseppunct}\relax
\EndOfBibitem
\bibitem[May and K{\"{u}}hn(2004)May, and K{\"{u}}hn]{May2004}
May,~V.; K{\"{u}}hn,~O. \emph{{Charge and Energy Transfer Dynamics in Molecular
  Systems, 3rd, Revised and Enlarged Edition}}; Wiley: Weinheim, 2004\relax
\mciteBstWouldAddEndPuncttrue
\mciteSetBstMidEndSepPunct{\mcitedefaultmidpunct}
{\mcitedefaultendpunct}{\mcitedefaultseppunct}\relax
\EndOfBibitem
\bibitem[Heyden \latin{et~al.}(2010)Heyden, Sun, Funkner, Mathias, Forbert,
  Havenith, and Marx]{marx:2010}
Heyden,~M.; Sun,~J.; Funkner,~S.; Mathias,~G.; Forbert,~H.; Havenith,~M.;
  Marx,~D. Dissecting the THz Spectrum of Liquid Water from First Principles
  Via Correlations in Time and Space. \emph{Proc. Natl. Acad. Sci.}
  \textbf{2010}, \emph{107}, 12068--12073\relax
\mciteBstWouldAddEndPuncttrue
\mciteSetBstMidEndSepPunct{\mcitedefaultmidpunct}
{\mcitedefaultendpunct}{\mcitedefaultseppunct}\relax
\EndOfBibitem
\bibitem[Moreau and Douh{\'e}ret(1974)Moreau, and Douh{\'e}ret]{moreau:1974}
Moreau,~C.; Douh{\'e}ret,~G. Solvatation Ionique Dans Les M{\'e}langes
  Eau-ac{\'e}tonitrile. Structure De Ceux-ci. \emph{J. Chim. Phys.}
  \textbf{1974}, \emph{71}, 1313--1321\relax
\mciteBstWouldAddEndPuncttrue
\mciteSetBstMidEndSepPunct{\mcitedefaultmidpunct}
{\mcitedefaultendpunct}{\mcitedefaultseppunct}\relax
\EndOfBibitem
\bibitem[Reimers and Hall(1999)Reimers, and Hall]{reimers:1999}
Reimers,~J.~R.; Hall,~L.~E. The Solvation of Acetonitrile. \emph{J. Am. Chem.
  Soc.} \textbf{1999}, \emph{121}, 3730--3744\relax
\mciteBstWouldAddEndPuncttrue
\mciteSetBstMidEndSepPunct{\mcitedefaultmidpunct}
{\mcitedefaultendpunct}{\mcitedefaultseppunct}\relax
\EndOfBibitem
\bibitem[Fouqueau \latin{et~al.}(2007)Fouqueau, Meuwly, and
  Bemish]{MM.acn:2007}
Fouqueau,~A.; Meuwly,~M.; Bemish,~R.~J. Adsorption of Acridine Orange at a C8,
  18/Water/Acetonitrile Interface. \emph{J. Phys. Chem. B} \textbf{2007},
  \emph{111}, 10208--10216\relax
\mciteBstWouldAddEndPuncttrue
\mciteSetBstMidEndSepPunct{\mcitedefaultmidpunct}
{\mcitedefaultendpunct}{\mcitedefaultseppunct}\relax
\EndOfBibitem
\bibitem[Besnard \latin{et~al.}(1992)Besnard, Caba{\c{c}}o, Strehle, and
  Yarwood]{besnard:1992}
Besnard,~M.; Caba{\c{c}}o,~M.~I.; Strehle,~F.; Yarwood,~J. Raman Spectroscopic
  Studies on the Dynamic and Equilibrium Processes in Binary Mixtures
  Containing Methanol and Acetonitrile. \emph{Chem. Phys.} \textbf{1992},
  \emph{163}, 103--114\relax
\mciteBstWouldAddEndPuncttrue
\mciteSetBstMidEndSepPunct{\mcitedefaultmidpunct}
{\mcitedefaultendpunct}{\mcitedefaultseppunct}\relax
\EndOfBibitem
\bibitem[Chou \latin{et~al.}(2012)Chou, Soper, Khodadadi, Curtis, Krueger,
  Cicerone, Fitch, and Shalaev]{chou:2012}
Chou,~S.~G.; Soper,~A.~K.; Khodadadi,~S.; Curtis,~J.~E.; Krueger,~S.;
  Cicerone,~M.~T.; Fitch,~A.~N.; Shalaev,~E.~Y. Pronounced Microheterogeneity
  in a Sorbitol--Water Mixture Observed through Variable Temperature Neutron
  Scattering. \emph{J. Phys. Chem. B} \textbf{2012}, \emph{116},
  4439--4447\relax
\mciteBstWouldAddEndPuncttrue
\mciteSetBstMidEndSepPunct{\mcitedefaultmidpunct}
{\mcitedefaultendpunct}{\mcitedefaultseppunct}\relax
\EndOfBibitem
\end{mcitethebibliography}


\providecommand{\latin}[1]{#1}
\makeatletter
\providecommand{\doi}
  {\begingroup\let\do\@makeother\dospecials
  \catcode`\{=1 \catcode`\}=2 \doi@aux}
\providecommand{\doi@aux}[1]{\endgroup\texttt{#1}}
\makeatother
\providecommand*\mcitethebibliography{\thebibliography}
\csname @ifundefined\endcsname{endmcitethebibliography}
  {\let\endmcitethebibliography\endthebibliography}{}
\begin{mcitethebibliography}{9}
\providecommand*\natexlab[1]{#1}
\providecommand*\mciteSetBstSublistMode[1]{}
\providecommand*\mciteSetBstMaxWidthForm[2]{}
\providecommand*\mciteBstWouldAddEndPuncttrue
  {\def\EndOfBibitem{\unskip.}}
\providecommand*\mciteBstWouldAddEndPunctfalse
  {\let\EndOfBibitem\relax}
\providecommand*\mciteSetBstMidEndSepPunct[3]{}
\providecommand*\mciteSetBstSublistLabelBeginEnd[3]{}
\providecommand*\EndOfBibitem{}
\mciteSetBstSublistMode{f}
\mciteSetBstMaxWidthForm{subitem}{(\alph{mcitesubitemcount})}
\mciteSetBstSublistLabelBeginEnd
  {\mcitemaxwidthsubitemform\space}
  {\relax}
  {\relax}

\bibitem[Vanommeslaeghe \latin{et~al.}(2010)Vanommeslaeghe, Hatcher, Acharya,
  Kundu, Zhong, Shim, Darian, Guvench, Lopes, Vorobyov, and
  Mackerell~Jr.]{CGenFF}
Vanommeslaeghe,~K.; Hatcher,~E.; Acharya,~C.; Kundu,~S.; Zhong,~S.; Shim,~J.;
  Darian,~E.; Guvench,~O.; Lopes,~P.; Vorobyov,~I.; Mackerell~Jr.,~A.~D. Charmm
  General Force Field: A Force Field for Drug-Like Molecules Compatible with
  the Charmm All-Atom Additive Biological Force Fields. \emph{J. Comp. Chem.}
  \textbf{2010}, \emph{31}, 671--690\relax
\mciteBstWouldAddEndPuncttrue
\mciteSetBstMidEndSepPunct{\mcitedefaultmidpunct}
{\mcitedefaultendpunct}{\mcitedefaultseppunct}\relax
\EndOfBibitem
\bibitem[Jorgensen \latin{et~al.}(1983)Jorgensen, Chandrasekhar, Madura, Impey,
  and Klein]{TIP3P-Jorgensen-1983}
Jorgensen,~W.~L.; Chandrasekhar,~J.; Madura,~J.~D.; Impey,~R.~W.; Klein,~M.~L.
  Comparison of Simple Potential Functions for Simulating Liquid Water.
  \emph{J. Chem. Phys.} \textbf{1983}, \emph{79}, 926--935\relax
\mciteBstWouldAddEndPuncttrue
\mciteSetBstMidEndSepPunct{\mcitedefaultmidpunct}
{\mcitedefaultendpunct}{\mcitedefaultseppunct}\relax
\EndOfBibitem
\bibitem[Beglov and Roux(1994)Beglov, and Roux]{benoit:1994}
Beglov,~D.; Roux,~B. Finite Representation of an Infinite Bulk System: Solvent
  Boundary Potential for Computer Simulations. \emph{J. Chem. Phys.}
  \textbf{1994}, \emph{100}, 9050--9063\relax
\mciteBstWouldAddEndPuncttrue
\mciteSetBstMidEndSepPunct{\mcitedefaultmidpunct}
{\mcitedefaultendpunct}{\mcitedefaultseppunct}\relax
\EndOfBibitem
\bibitem[Bian \latin{et~al.}(2013)Bian, Chen, Zhang, Li, Wen, Zhuang, and
  Zheng]{bian2013cation}
Bian,~H.; Chen,~H.; Zhang,~Q.; Li,~J.; Wen,~X.; Zhuang,~W.; Zheng,~J. Cation
  Effects on Rotational Dynamics of Anions and Water Molecules in Alkali
  (Li$^+$, Na$^+$, K$^+$, Cs$^+$) Thiocyanate (SCN$^-$) Aqueous Solutions.
  \emph{J. Phys. Chem. B} \textbf{2013}, \emph{117}, 7972--7984\relax
\mciteBstWouldAddEndPuncttrue
\mciteSetBstMidEndSepPunct{\mcitedefaultmidpunct}
{\mcitedefaultendpunct}{\mcitedefaultseppunct}\relax
\EndOfBibitem
\bibitem[Berendsen \latin{et~al.}(1987)Berendsen, Grigera, and
  Straatsma]{straatsma:1987spce}
Berendsen,~H. J.~C.; Grigera,~J.~R.; Straatsma,~T.~P. The Missing Term in
  Effective Pair Potentials. \emph{J. Phys. Chem.} \textbf{1987}, \emph{91},
  6269--6271\relax
\mciteBstWouldAddEndPuncttrue
\mciteSetBstMidEndSepPunct{\mcitedefaultmidpunct}
{\mcitedefaultendpunct}{\mcitedefaultseppunct}\relax
\EndOfBibitem
\bibitem[Liu \latin{et~al.}(2013)Liu, Wei, Zhao, and Wang]{wang:2013}
Liu,~B.; Wei,~F.; Zhao,~J.; Wang,~Y. Characterization of Amide–thiocyanates
  Eutectic Ionic Liquids and their Application in SO$_2$ Absorption. \emph{RSC
  Adv.} \textbf{2013}, \emph{3}, 2470--2476\relax
\mciteBstWouldAddEndPuncttrue
\mciteSetBstMidEndSepPunct{\mcitedefaultmidpunct}
{\mcitedefaultendpunct}{\mcitedefaultseppunct}\relax
\EndOfBibitem
\bibitem[Mitchell \latin{et~al.}(1992)Mitchell, Butler, and
  Albright]{albright:1992}
Mitchell,~J.~P.; Butler,~J.~B.; Albright,~J.~G. Measurement of Mutual Diffusion
  Coefficients, Densities, Viscosities, and Osmotic Coefficients for the System
  KSCN-H$_2$O at 25°C. \emph{J Solution Chem} \textbf{1992}, \emph{21},
  1115--1129\relax
\mciteBstWouldAddEndPuncttrue
\mciteSetBstMidEndSepPunct{\mcitedefaultmidpunct}
{\mcitedefaultendpunct}{\mcitedefaultseppunct}\relax
\EndOfBibitem
\bibitem[Liu \latin{et~al.}(2013)Liu, Zhao, and Wei]{liu:2013}
Liu,~B.; Zhao,~J.; Wei,~F. Effects of Water on the Properties of Acetamide-KSCN
  Eutectic Ionic Liquids at Several Temperatures. \emph{J. Mol. Liq.}
  \textbf{2013}, \emph{187}, 309--313\relax
\mciteBstWouldAddEndPuncttrue
\mciteSetBstMidEndSepPunct{\mcitedefaultmidpunct}
{\mcitedefaultendpunct}{\mcitedefaultseppunct}\relax
\EndOfBibitem
\end{mcitethebibliography}

\end{document}


\clearpage

\section*{Force Field Parametrization}

\begin{table}
\caption{Bonded and non-bonded parameters.}
\label{sitab:params}
\begin{tabular}{c|ccc}
\hline\hline
\textbf{Residues} & \multicolumn{3}{c}{Parameters} \\
\hline \hline
\textbf{Acetamide} & \multicolumn{3}{c}{CGenFF\cite{CGenFF}} \\
\hline \hline
\textbf{Water} & \multicolumn{3}{c}{TIP3P\cite{TIP3P-Jorgensen-1983}} \\
\hline \hline
\textbf{K$^+$} & \multicolumn{3}{c}{} \\
\hline
Non-bonded & $\epsilon$ (kcal/mol)& $r_\mathrm{min}$ (\AA) & $q$ (e)\\
K$^+$\cite{benoit:1994} & $0.087$ & $3.5275$ & $+1$ \\
\hline \hline 
\textbf{SCN$^-$} & \multicolumn{3}{c}{} \\
\hline
Bond (Morse Potential) & $r_\mathrm{eq}$ (\AA) & $\beta$ (\AA$^{-1}$)& $D_0$ (kcal/mol)\\
C-N * & $1.21706$ & $1.61348$ & $373.811$ \\
S-C * & $1.66867$ & $1.69773$ & $124.360$ \\
\hline
Angle (Harmonic Potential) & $k_\theta$ (kcal/mol/rad$^2$) & $\theta_\mathrm{min}$ ($^\circ$)& \\
S-C-N & $21.20$ & $1.61348$ & \\
\hline 
Non-bonded & $\epsilon$ (kcal/mol)& $r_\mathrm{min}$ (\AA) & $q$ (e) \\
S & $0.45$ & $4.00$ & $-0.183$ \\
C & $0.18$ & $3.58$ & $-0.362$ \\
N & $0.18$ & $3.74$ & $-0.455$ \\
\hline 
Atomic Multipoles ** & S & C & N \\
$Q_{00}$  & $-0.183$ & $-0.362$ & $-0.455$ \\
$Q_{10}$  & $1.179$ & $0.163$ & $0.319$ \\
$Q_{11c}$ & $0.0$ & $0.0$ & $0.0$ \\
$Q_{11s}$ & $0.0$ & $0.0$ & $0.0$ \\
$Q_{20}$  & $-1.310$ & $-0.929$ & $-3.114$ \\
$Q_{21c}$ & $0.0$ & $0.0$ & $0.0$ \\
$Q_{21s}$ & $0.0$ & $0.0$ & $0.0$ \\
$Q_{22c}$ & $0.0$ & $0.0$ & $0.0$ \\
$Q_{22s}$ & $0.0$ & $0.0$ & $0.0$ \\
\hline \hline
\multicolumn{4}{l}{* MP2/cc-aug-pVDZ}\\
\multicolumn{4}{l}{** within Local Reference Axis System and in atomic units - see MTPL documentation:} \\
\multicolumn{4}{l}{\url{www.charmm.org/archive/charmm/documentation/by-version/c45b1/mtpl.html}}
\end{tabular}
\end{table}

\clearpage

\section*{System compositions}

\begin{table}
\caption{Molar fraction and number of molecules used on the MD
  simulations for each component of the systems at a given
  water/acetamide mixing ratio.}
\label{sitab:composition}
\begin{tabular}{c|ccc||ccc}
\hline\hline
& \multicolumn{3}{c||}{Molar Fraction} & \multicolumn{3}{c}{Number of Molecules} \\ \hline
Mixing Ratio & \multicolumn{1}{c|}{H$_{2}$O}   & \multicolumn{1}{c|}{Acetamide} & KSCN  & \multicolumn{1}{c|}{H$_{2}$O} & \multicolumn{1}{c|}{Acetamide} & KSCN \\
\hline \hline
100                            & \multicolumn{1}{c|}{0.901} & \multicolumn{1}{c|}{0.000}     & 0.099 & \multicolumn{1}{c|}{685} & \multicolumn{1}{c|}{0}         & 75   \\
95                             & \multicolumn{1}{c|}{0.844} & \multicolumn{1}{c|}{0.044}     & 0.111 & \multicolumn{1}{c|}{570} & \multicolumn{1}{c|}{30}        & 75   \\
90                             & \multicolumn{1}{c|}{0.791} & \multicolumn{1}{c|}{0.092}     & 0.118 & \multicolumn{1}{c|}{506} & \multicolumn{1}{c|}{59}        & 75   \\
80                             & \multicolumn{1}{c|}{0.686} & \multicolumn{1}{c|}{0.179}     & 0.135 & \multicolumn{1}{c|}{381} & \multicolumn{1}{c|}{100}       & 75   \\
70                             & \multicolumn{1}{c|}{0.586} & \multicolumn{1}{c|}{0.263}     & 0.152 & \multicolumn{1}{c|}{290} & \multicolumn{1}{c|}{130}       & 75   \\
60                             & \multicolumn{1}{c|}{0.490} & \multicolumn{1}{c|}{0.342}     & 0.168 & \multicolumn{1}{c|}{219} & \multicolumn{1}{c|}{153}       & 75   \\
50                             & \multicolumn{1}{c|}{0.399} & \multicolumn{1}{c|}{0.418}     & 0.183 & \multicolumn{1}{c|}{164} & \multicolumn{1}{c|}{171}       & 75   \\
40                             & \multicolumn{1}{c|}{0.312} & \multicolumn{1}{c|}{0.490}     & 0.198 & \multicolumn{1}{c|}{119} & \multicolumn{1}{c|}{186}       & 75   \\
30                             & \multicolumn{1}{c|}{0.229} & \multicolumn{1}{c|}{0.559}     & 0.212 & \multicolumn{1}{c|}{81}  & \multicolumn{1}{c|}{198}       & 75   \\
20                             & \multicolumn{1}{c|}{0.149} & \multicolumn{1}{c|}{0.626}     & 0.225 & \multicolumn{1}{c|}{50}  & \multicolumn{1}{c|}{209}       & 75   \\
10                             & \multicolumn{1}{c|}{0.073} & \multicolumn{1}{c|}{0.689}     & 0.238 & \multicolumn{1}{c|}{23}  & \multicolumn{1}{c|}{217}       & 75   \\
0                              & \multicolumn{1}{c|}{0.000} & \multicolumn{1}{c|}{0.750}     & 0.250 & \multicolumn{1}{c|}{0}   & \multicolumn{1}{c|}{225}       & 75   \\ \hline \hline
\end{tabular}
\end{table}

\clearpage

\section*{Densities}

\begin{figure}
\centering
\includegraphics[width=0.75\textwidth]{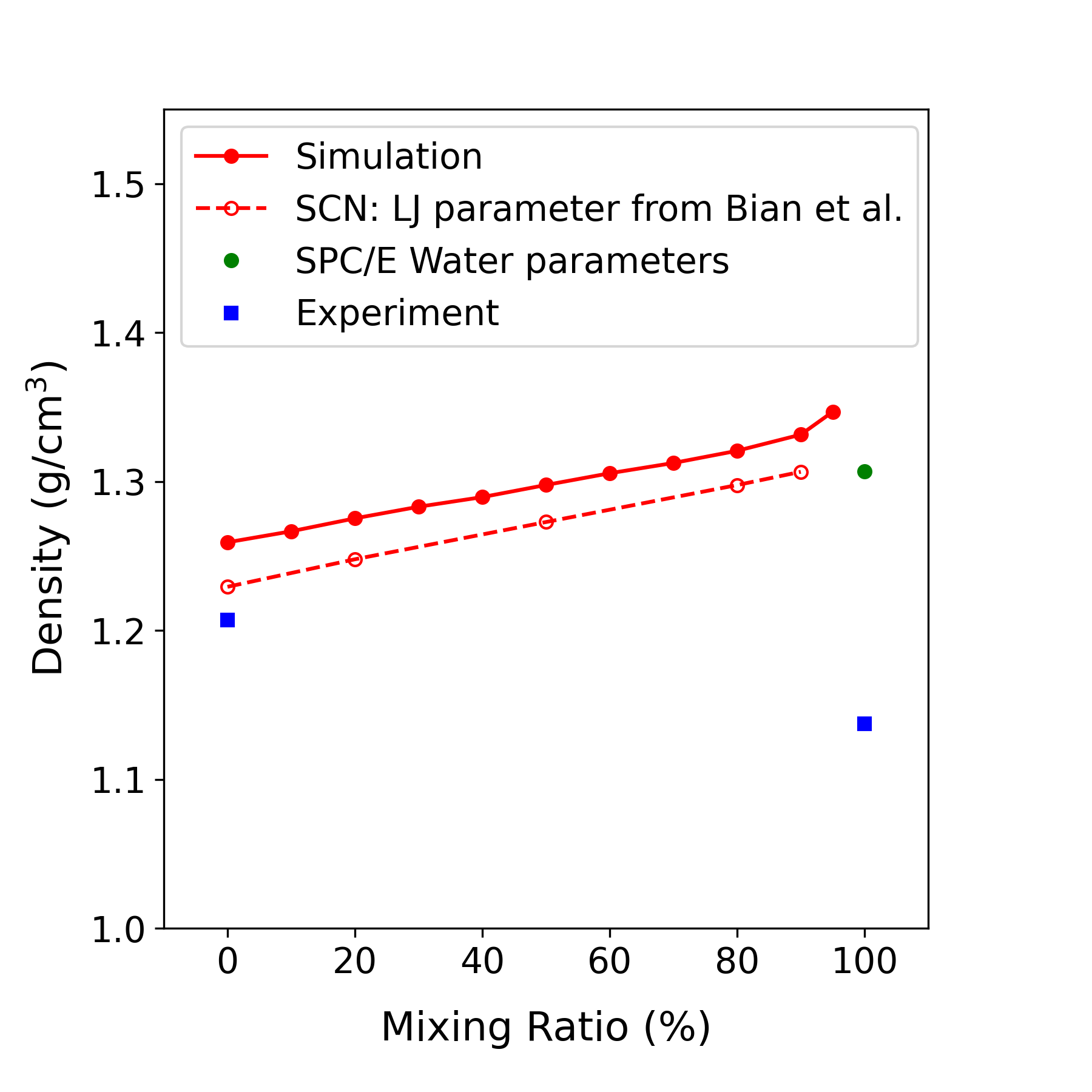}
\caption{Average density of the simulation systems for different
  mixing rations in the \emph{NPT} at 300\,K (red full circles).
  Average densities from MD simulation with Lennard-Jones parameters
  for SCN$^-$ from Bian et al.\cite{bian2013cation} are shown with
  open red circles and from MD simulation with the SPC/E water
  parametrization for the 100\% mixture are shown as green
  circle.\cite{straatsma:1987spce} Experimental values at 298\,K for
  KSCN in acetamide\cite{wang:2013} and in water\cite{albright:1992}
  are shown with blue squares. The experimental density for KSCN in
  water is linearly interpolated between the closest reference values
  to fit the molality of KSCN in the simulated system at 95\% mixing
  ratio ($3.77$ mol/kg).}
\label{sifig:1}
\end{figure}

\clearpage
\section*{Frequency-Frequency Correlation Functions}

\begin{figure}
\centering
\includegraphics[width=0.75\textwidth]{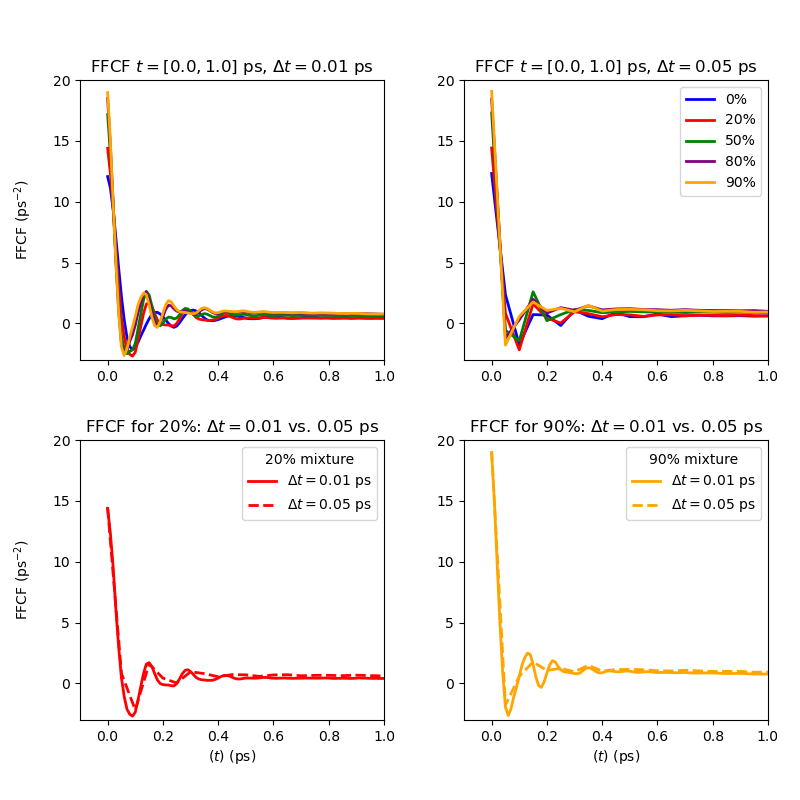}
\caption{Averaged frequency-frequency correlation function (FFCF) of
  the INM frequencies of of all C-N vibrations in the SCN$^-$ anion in
  different mixing ratios of acetamide and water. The FFCF are
  computed from INM frequency sequences of length $t=250$~ps with time
  steps of $t=0.01$~ps (top left panel) and of length $t=1000$~ps with
  time steps of $t=0.05$~ps (top right panel). Bottom left and right
  panels directly compare the FFCFs for mixture 20\% and 90\% with
  different time steps, respectively.}
\label{sifig:2}
\end{figure}

\clearpage

\section*{THz Spectra}

\begin{figure}
\centering
\includegraphics[width=0.75\textwidth]{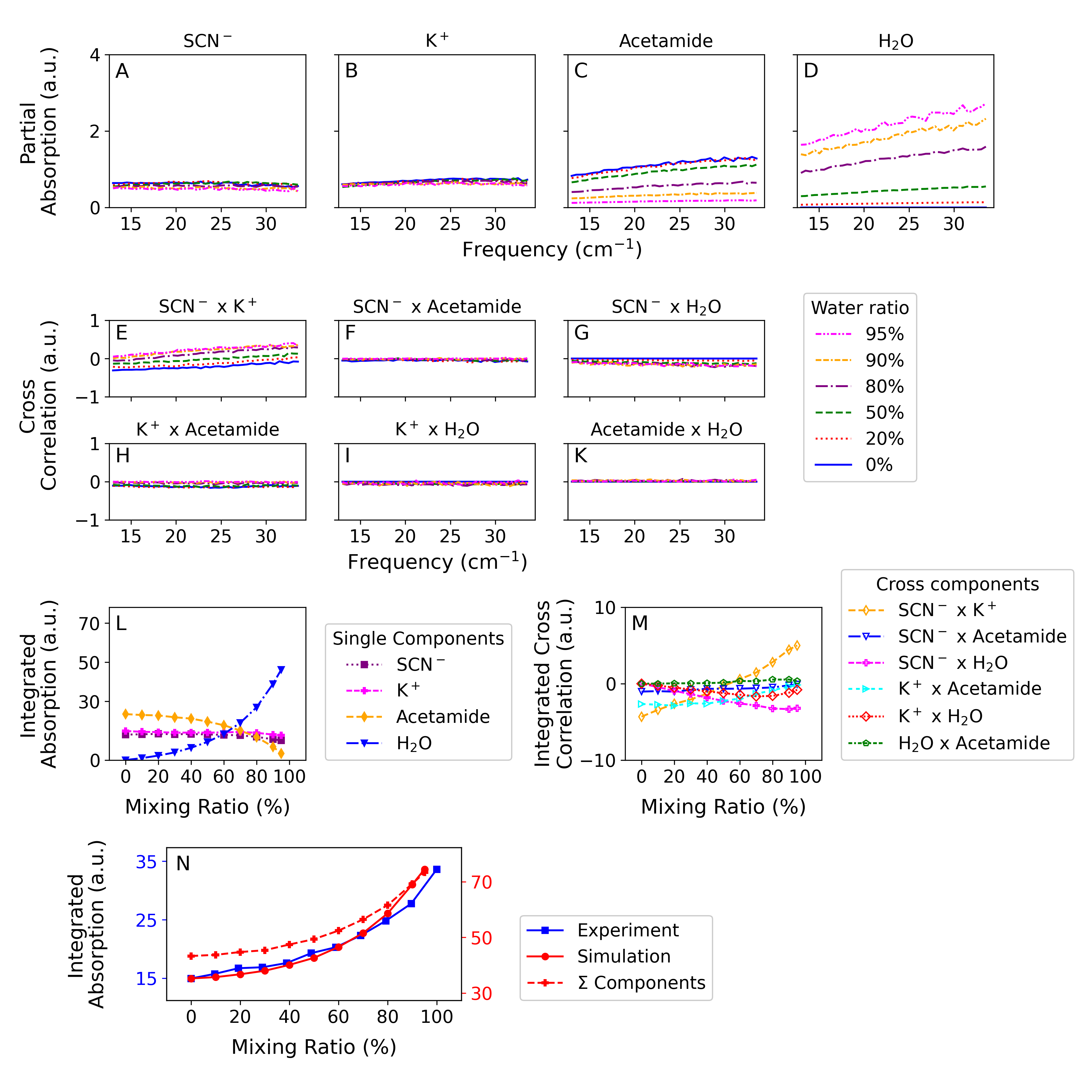}
\caption{Components of the THz absorption spectra from the Fourier
  transform of the (A-D) auto-correlation of the dipole sequences of
  all (A) SCN$^-$ anions, (B) K$^+$ cations, (C) H$_2$O and (D)
  acetamide molecules.  Panels E-K shows the Fourier transform of the
  cross-correlation functions of the 6 possible dipole sequence
  combinations.  The integrated intensity of the single component THz
  absorption spectra is given in panel L and of the cross correlation
  functions in panel M. Panel N compares the integrated absorption of
  the experimental THz spectra (blue line) for different mixtures with
  the simulated spectra from the total dipole sequence (red line) and
  the sum of the single components in panel L and cross components in
  panel M (red dashed line).}
\label{sifig:3}
\end{figure}

\clearpage

\section*{Vibrational Coupling}

\begin{figure}
\centering
\includegraphics[width=0.75\textwidth]{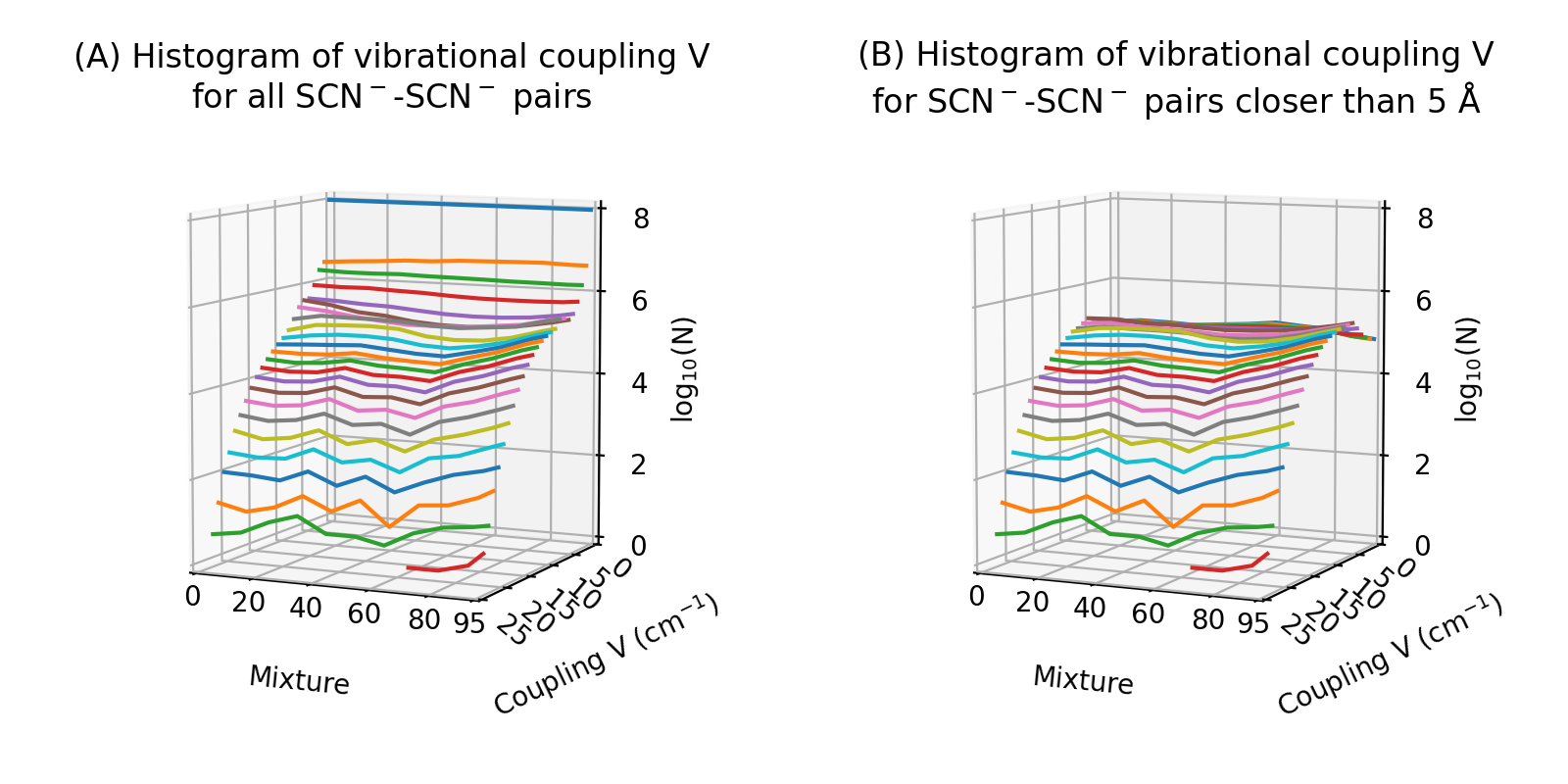}
\caption{Vibrational transition coupling strength $V$ (A) between all
  SCN$^-$-SCN$^-$ ion pairs and (B) for pair distances shorter than
  5\,\AA. Each curve shows the change in amplitude for different
  mixing ratios of the histogram of ion pair number vs. transition
  coupling with bin size of 1\,cm$^{-1}$. The coupling is computed for
  the same data selection used for the computation of the 2D~IR
  spectra ($\sim 10^8$ ion pairs per mixture). The line shapes of the
  curves for medium and high coupling strengths are unaffected by the
  exclusion of ion pairs with distances larger than 5\,\AA.}
\label{sifig:4}
\end{figure}

\clearpage

\section*{Radial Distribution Function at Experimental Density}
The computed system density of the \emph{NPT} ensemble at 300~K and
normal pressure is almost linearly increasing from 1.26~g/cm$^3$ (0\%)
to 1.35~g/cm$^3$ (95\%). Experimentally obtained reference densities
yield a decrease from 0\% to 100\% water content from 1.207~g/cm$^3$
to 1.14~g/cm$^3$.\cite{wang:2013, albright:1992} Experimental
densities for a KSCN:acetamide DES mixed in different ratios with
water are reported in reference~\citenum{liu:2013}, however, in their
work the KSCN concentration decreases with increasing water ratio.
Additional simulation for a subset of mixing ratios were performed
using Lennard-Jones Parameters for SCN$^-$ from the work of Bian et
al.\cite{bian2013cation} (open red circles). Additionally, one MD
simulation for the 100\% mixture using the SPC/E water model was
performed that yields an average density of
1.31~g/cm$^3$.\cite{straatsma:1987spce} One final simulation was
carried out for KSCN in 100 \% water at the experimentally determined
density of $1.14$~g/cm$^3$ to determine the sensitivity of the
SCN$^-$/SCN$^-$ ordering on the density of the simulated system, see
Figure \ref{sifig:2}. It is found that the radial distribution
function $g_{\rm SCN^{-}-SCN^{-}}(r_{\rm CC})$ only marginally changes
if the density increases from $1.14$~g/cm$^3$ to $1.35$~g/cm$^3$.\\

\begin{figure}
\centering
\includegraphics[width=0.75\textwidth]{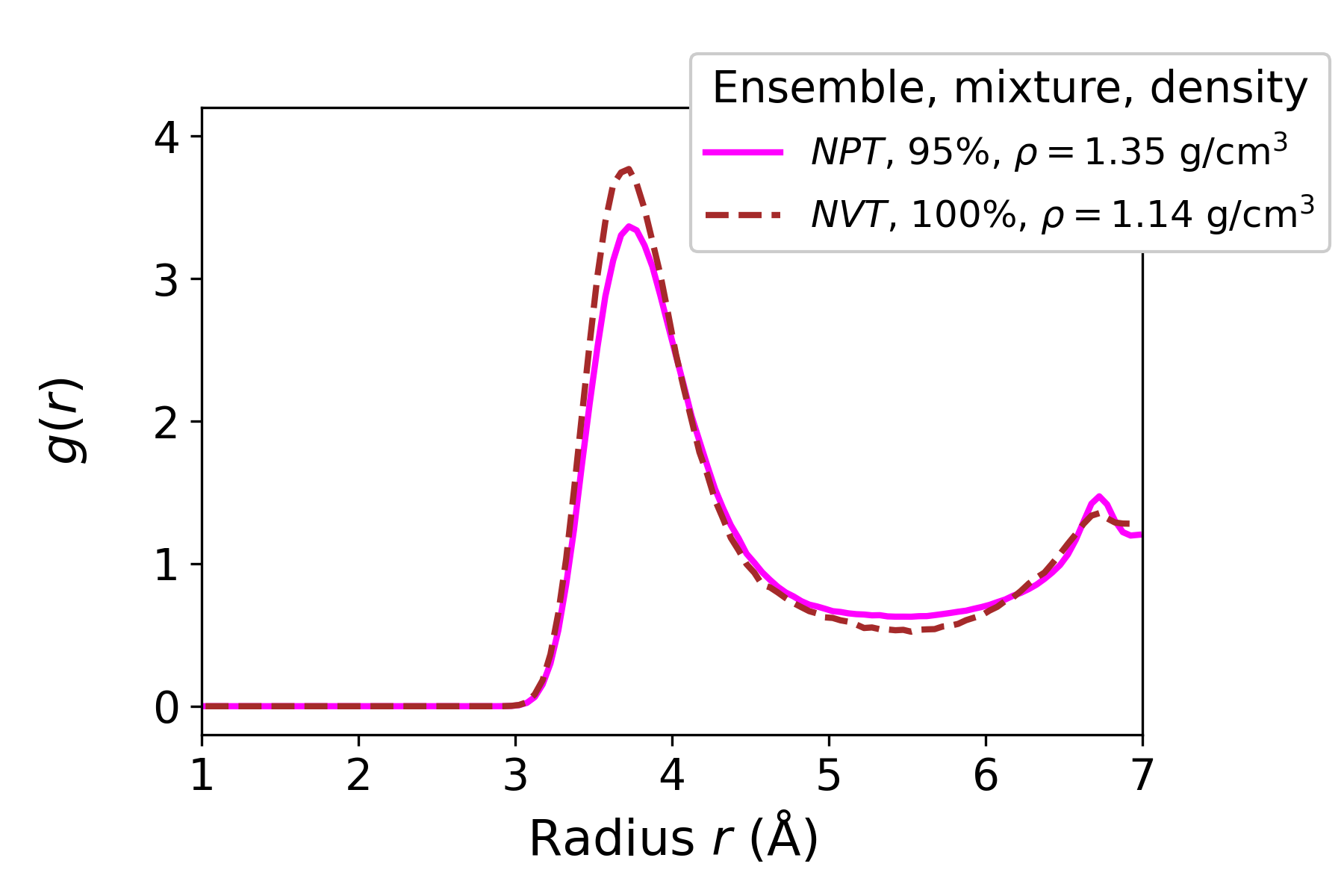}
\caption{Radial Distribution function $g_{\rm SCN^{-}-SCN^{-}}(r_{\rm
    CC})$ between the C atoms of SCN$^-$ anions from MD simulations
  with parametrization used throughout this work (see Table
  \ref{sitab:params}) of a \emph{NVT} ensemble of KSCN in water with a
  respective box size to match the experimentally observed density of
  $1.14$~g/cm$^3$ (dashed brown line) with KSCN molality of
  $3.77$~mol/kg. The radial distribution function in the \emph{NPT}
  ensemble of KSCN in a 95\% mixture of water and acetamide with
  average density of $1.35$~g/cm$^3$ is shown by the solid magenta
  line, see also Figure 4.}
\label{sifig:5}
\end{figure}

\clearpage
\bibliography{refs}

\clearpage